\documentclass[onecolumn,preprint]{aastex61}

\begin{document}


\title{A panchromatic view of the Bulge globular cluster NGC 6569\footnote{\footnotesize Based on
		observations obtained at the Gemini Observatory, which is operated
		by the Association of Universities for Research in Astronomy,
		Inc., under a cooperative agreement with the NSF on behalf of the
		Gemini partnership: the National Science Foundation (United
		States), the National Research Council (Canada), CONICYT (Chile),
		the Australian Research Council (Australia), Minist\'{e}rio da
		Ci\^{e}ncia, Tecnologia e Inova\c{c}\~{a}o (Brazil) and Ministerio
		de Ciencia, Tecnolog\'{i}a e Innovaci\'{o}n Productiva
		(Argentina). Based on observations (GO 15232, PI: Ferraro) with the NASA/ESA 
		\textit{Hubble Space Telescope}, obtained at the Space Telescope Science 
		Institute, which is operated by AURA, Inc., under NASA contract 5-26555.}}

%% Use \author, \affil, plus the \and command to format author and affiliation 
%% information.  If done correctly the peer review system will be able to
%% automatically put the author and affiliation information from the manuscript
%% and save the corresponding author the trouble of entering it by hand.
%%
%% The \affil should be used to document primary affiliations and the
%% \altaffil should be used for secondary affiliations, titles, or email.

%% Authors with the same affiliation can be grouped in a single
\author{S. Saracino} \affiliation{Dipartimento di Fisica e Astronomia,
	Universit\`a di Bologna, Via Gobetti 93/2, I-40129 Bologna, Italy}
\affiliation{INAF - Osservatorio di Astrofisica e Scienza dello Spazio, via Gobetti
	93/3, I-40129 Bologna, Italy} \correspondingauthor{Sara Saracino} \email{sara.saracino@unibo.it}

\author{E. Dalessandro} \affiliation{INAF - Osservatorio di Astrofisica e Scienza dello Spazio, via Gobetti
	93/3, I-40129 Bologna, Italy}

\author{F. R. Ferraro} \affiliation{Dipartimento di Fisica e
	Astronomia, Universit\`a di Bologna, Via Gobetti 93/2, I-40129
	Bologna, Italy}
\affiliation{INAF - Osservatorio di Astrofisica e Scienza dello Spazio, via Gobetti
	93/3, I-40129 Bologna, Italy}

\author{B. Lanzoni} \affiliation{Dipartimento di Fisica e Astronomia,
	Universit\`a di Bologna, Via Gobetti 93/2, I-40129 Bologna, Italy}
\affiliation{INAF - Osservatorio di Astrofisica e Scienza dello Spazio, via Gobetti
	93/3, I-40129 Bologna, Italy}

\author{D. Geisler} \affiliation{Departamento de Astronom\'ia,
	Universidad de Concepci\'on, Casilla 160-C, Concepci\'on, Chile}
\affiliation{Instituto de Investigaci\'on Multidisciplinario en Ciencia y Tecnolog\'ia Universidad de La Serena, Chile}
\affiliation{Departamento de F\'isica y Astronom\'ia, Facultad de Ciencias, Universidad de La Serena, Av. Juan Cisternas 1200, La Serena, Chile}

\author{R. E. Cohen} \affiliation{Space Telescope Science Institute,
	3700 San Martin Drive, Baltimore, MD 21218, USA}

\author{A. Bellini} \affiliation{Space Telescope Science Institute,
	3700 San Martin Drive, Baltimore, MD 21218, USA}

\author{E. Vesperini} \affiliation{Department of Astronomy, Indiana 
	University, Bloomington, IN 47405, USA}

\author{M. Salaris} \affiliation{Astrophysics Research Institute,
	Liverpool John Moores University, 146 Brownlow Hill, Liverpool L3
	5RF, UK}

\author{S. Cassisi} \affiliation{INAF - Osservatorio Astronomico d'Abruzzo, 
	sn., I-64100 Teramo, Italy}

\author{A. Pietrinferni} \affiliation{INAF - Osservatorio Astronomico d'Abruzzo, 
	sn., I-64100 Teramo, Italy}

\author{L. Origlia} \affiliation{INAF - Osservatorio di Astrofisica e Scienza dello Spazio, via Gobetti
	93/3, I-40129 Bologna, Italy}

\author{F. Mauro} \affiliation{Instituto de Astronom\'ia, Universidad
	Cat\'olica del Norte, Av. Angamos 0610, Antofagasta, Chile}

\author{S. Villanova} \affiliation{Departamento de Astronom\'ia,
	Universidad de Concepci\'on, Casilla 160-C, Concepci\'on, Chile}

\author{C. Moni Bidin} \affiliation{Instituto de Astronom\'ia,
	Universidad Cat\'olica del Norte, Av. Angamos 0610, Antofagasta,
	Chile}

\begin{abstract}

We used high-resolution optical {\it HST}/WFC3 and multi-conjugate adaptive optics assisted GEMINI GeMS/GSAOI observations in the near-infrared to investigate the physical properties of the globular cluster NGC 6569 in the Galactic bulge. We have obtained the deepest purely NIR color-magnitude diagram published so far for this cluster using ground-based observations, reaching $K_{s}$ $\approx$ 21.0 mag (two magnitudes below the main-sequence turn-off point). By combining the two datasets secured at two different epochs, we determined relative proper motions for a large sample of individual stars in the center of NGC 6569, allowing a robust selection of cluster member stars. Our proper motion analysis solidly demonstrates that, despite its relatively high metal content, NGC 6569 hosts some blue horizontal branch stars. A differential reddening map has been derived in the direction of the system, revealing a maximum color excess variation of about $\delta E(B-V)$ $\sim$ 0.12 mag in the available field of view. The absolute age of NGC 6569 has been determined for the first time. In agreement with the other few bulge globular clusters with available age estimates, NGC 6569 turns out to be old, with an age of about 12.8 Gyr, and a typical uncertainty of 0.8-1.0 Gyr. %This is a somehow pioneering and promising result in the context of using the next generation of large ground-based telescopes equipped with AO modules, like the ESO/ELT and the TMT, to perform accurate astrometry in dense stellar fields.
\end{abstract}

\keywords{globular clusters: individual (NGC 6569) --- techniques: photometry, proper motion --- 
instrumentation: adaptive optics}



\section{Introduction} \label{sec:intro}
The Galactic bulge is one of the most inaccessible regions of the Milky Way and its structure, formation and evolution are still the subject of intense debate within the astronomical community (see for example \citealp{rich1998,ness2013,origlia2014,zoccali2016}). Many studies have shown that globular clusters (GCs) orbiting the bulge are key tools to trace the properties of the bulge stellar population in terms of kinematics, chemical abundances and age \citep{bica2006,valenti2007,barbuy2018}. Moreover, metal-rich bulge GCs represent the ideal local templates for studying the stellar content of extra-Galactic unresolved bulges and elliptical galaxies.
Unfortunately, because of observational limitations mainly related to the large extinction and stellar density in the direction of the bulge, its GCs have been systematically excluded from large surveys and remain poorly investigated.
The advent of high-resolution near-infrared (NIR) imagers equipped with Multi Conjugate Adaptive Optics (MCAO) facilities mounted at ground-based 8-10 m class telescopes opened a new line of investigation, arousing the interest of astronomers in the Galactic bulge and its bulge GC system \citep[e.g.][see also a complementary study based on deep NIR observations with the {\it Hubble Space Telescope (HST)} recently carried out by \citealt{cohen2018}]{ferraro2009a,saracino2015,saracino2016}.

Moreover, ground-based diffraction-limited observations provide very high spatial resolution (higher than {\it HST}, due to the larger apertures of the 8m-class telescopes) that they can be used for accurate proper motion (PM) studies. However, while the exceptional astrometric performance of {\it HST} has been extensively exploited during the last 20 years, allowing to investigate the kinematics and dynamics of a large sample of stellar systems (see e.g. \citealt{mclaughlin2006}; \citealt{anderson2010}; \citealt{mcnamara2012}; \citealt{bellini2014}; \citealt{watkins2015}; \citealt{libralato2018}), only a few PM studies using MCAO observations exist in the literature \citep{ortolani2011,massari2016,fritz2017,monty18}. %This is mostly due to the paucity oseveral distortion issues affecting MCAO data which are very difficult to model and remove. %It is therefore now essential to devote all the efforts to bring out the best from these systems.

The work presented here is focused on the GC NGC 6569 and it is part of an extended project aimed at characterizing the stellar populations of a sample of highly extincted stellar systems in the Galactic bulge (see the extensive work done in Terzan 5; \citealt{ferraro2009a,ferraro2015,ferraro2016}, \citealt{lanzoni2010}, \citealt{origlia2011,origlia2013}, \citealt{massari2012,massari2014a,massari2014b}) orbiting the innermost regions of our Galaxy. %The GC NGC 6569 is particularly suitable to be used as test case for the first multi-epoch astrometric study performed using both ground-based MCAO and space observations, without any previous star membership information.
 
NGC 6569 is a moderately compact globular (with a concentration parameter c = 1.27, \citealt{trager1995}) located in the Sagittarius region (l=$0.48^{\circ}$, b=$-6.68^{\circ}$; \citealt{harris1996}, 2010 edition), at a distance of only 3 kpc from the Galactic center \citep{harris1996}. The cluster is projected towards the dark nebula Barnard 305 \citep{barnard1927} and it is therefore highly reddened, with an average color excess $E(B-V)$ = 0.53 \citep{ortolani2001}. 
The first optical (V, V-I) color-magnitude diagram (CMD) of NGC 6569 was obtained \citep{ortolani2001} by using data from the 1.5 m Danish telescope at ESO La Silla. These data sample only the brightest portion of the CMD down to the red horizontal branch (HB). Deeper CMDs have been obtained in the optical by \cite{piotto2002} and in the NIR by \cite{valenti2005} using the {\it HST} and the New Technology Telescope (NTT) at ESO La Silla, respectively. However, these observations are unable to properly characterize the main-sequence turn-off (MS-TO) region of the cluster. A more recent photometric analysis of the cluster was performed by \citet[][see also \citealp{cohen2017}]{mauro2012} using $J$, $H$ and $K_{s}$ data from the Vista Variables in the Via Lactea (VVV) survey. Interestingly, the authors found evidence of the presence of two red HBs separated by $\approx$ 0.1 mag in the $K_{s}$ band. In a very recent work performed using the IR channel of the {\it HST}/WFC3, \cite{cohen2018} imaged the MS of NGC 6569 also in the NIR, sampling a few magnitudes below the MS-TO and highlighting the presence of an extension of the HB blueward of the red clump. 

From low-resolution spectroscopy, \cite{zinn1984} measured a metallicity [Fe/H]= $-0.8$ for the cluster. A similar value ([Fe/H]=$-0.79 \pm 0.02$) was obtained later from high-resolution IR spectra of six stars by \cite{valenti2011}, with an average $\alpha$-elements enhancement of +0.43 $\pm$ 0.02. The most recent determination has been published by \cite{johnson2018}, who studied a sample of 19 stars with the Magellan-M2FS and the VLT-FLAMES spectrographs and found a mean [Fe/H] = $-0.87$, with [$\alpha$/Fe] $\sim$ + 0.4.

One of the most surprising characteristics of NGC 6569, given its relatively high metallicity, is the presence of a quite populated family of RR Lyrae stars. The first evidence dates back to \cite{rosino1962}, who identified eight variable stars, of which six lie within the tidal radius of the cluster. Later on, \cite{hazenliller1984, hazenliller1985} and \cite{kunder2015} discovered a sizeable population of 27 RR Lyrae variables within 8 arcmin from the cluster center. %The presence of a significant number of these objects suggests that a non negligible fraction of HB stars crosses the instability strip. 
The proposed explanation for the large population of RR Lyrae was a possibly older age of NGC 6569 with respect to other GCs of the same metallicity \citep{zinn1980}. However, the age of the cluster has never been estimated.\\
In this work we present the first determinations of {\it(1)} relative PMs for a large sample of individual stars in the cluster core based on HST and GEMINI observations, allowing a solid separation between cluster members and field contaminants, {\it(2)} the differential reddening map in the direction of the system, and {\it(3)} the absolute age of the cluster. In Section 2 we describe the NIR and optical observations used in this work and the adopted data analysis. Section 3 presents the derived optical-NIR CMDs of NGC 6569, with a discussion of its main stellar populations. The PM analysis is described in Section 4. Section 5 is focused on the determination of the differential extinction in the direction of the cluster. In Section 6 we derive the distance modulus of NGC 6569 by means of two independent methods. The age of the cluster is determined in Section 7, by applying the MS-fitting method to different sets of theoretical models. In Section 8 we present our summary and conclusions.      

\section{Observations and data analysis} \label{sec:style}
\subsection{NIR and optical datasets}
\begin{figure}[ht!]
	\figurenum{1}
	\plotone{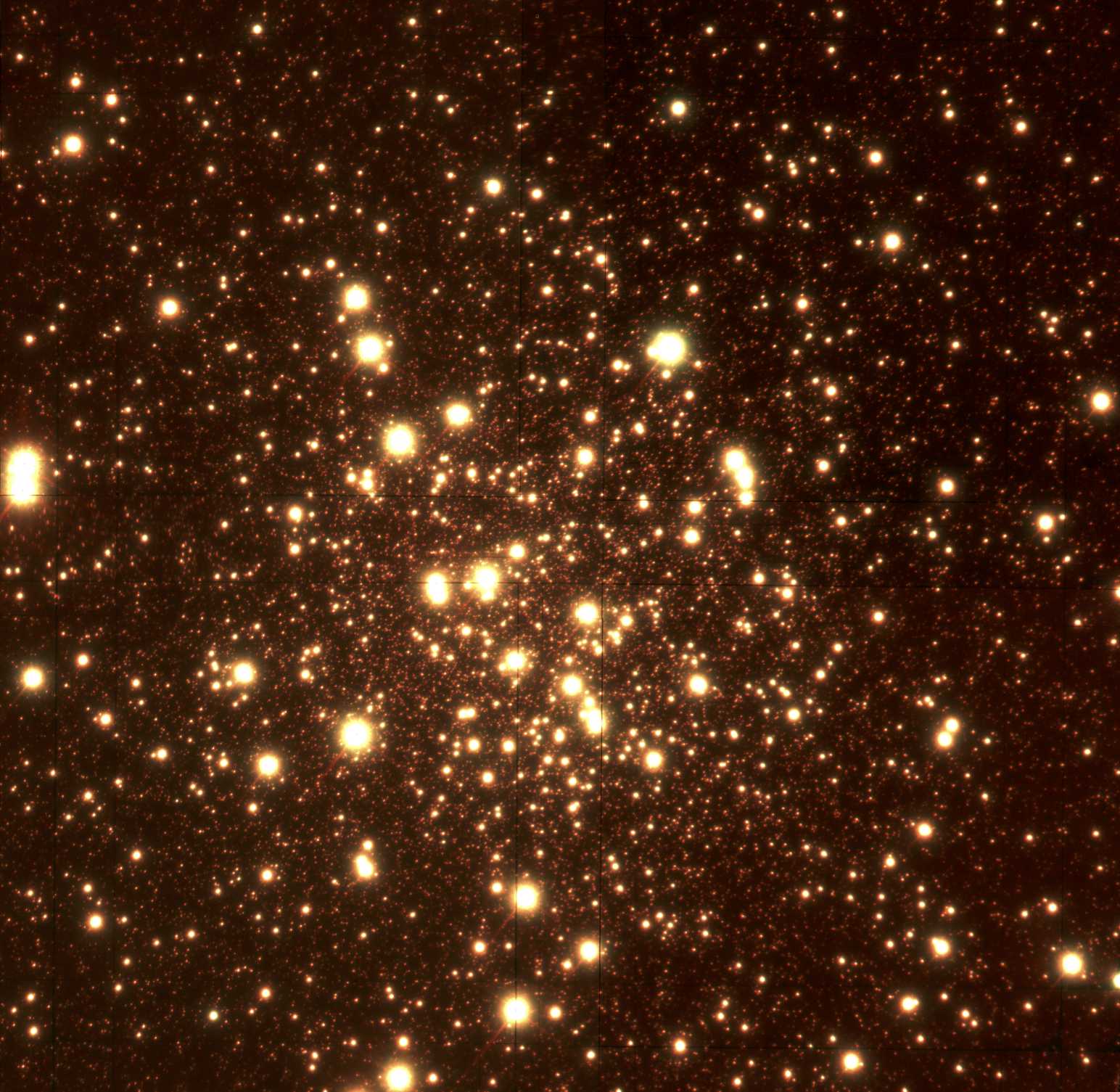}
	\caption{False-color image of NGC 6569 obtained by combining GEMINI observations in the near-IR J and $K_{s}$ bands. North is up; east is on the left. The FOV is 92" x 92".\label{fig:6569fig0}}
\end{figure}

The NIR photometric data used in the present study consist of a set of high resolution images obtained in May 2013 with the Gemini South Adaptive Optics Imager (GSAOI) assisted by GeMS at the 8 m Gemini South Telescope, as part of the GEMINI program GS-2013-Q-23, PI: D. Geisler. The camera GSAOI is characterized by four chips and it covers a total FOV of $85" \times 85"$ on the sky, with a high angular resolution (0.02"/pixel, \citealp{neichel2014}). GeMS uses a constellation of five laser guide stars plus three natural guide stars, to compensate the distortions due to the turbulence of the Earth's atmosphere. We sampled the central region of NGC 6569 with a mosaic of multiple exposures, 14 in both $J$ and $K_{s}$, with exposure time $t_{exp}$ = 30 sec each. A dither pattern of a few arcseconds has been applied in both filters to recover the gaps among the chips. In Figure \ref{fig:6569fig0} we show a two-color image of NGC 6569, obtained by combining GEMINI J and $K_{s}$-band observations. 

\begin{figure}[ht!]
	\figurenum{2}
	\plotone{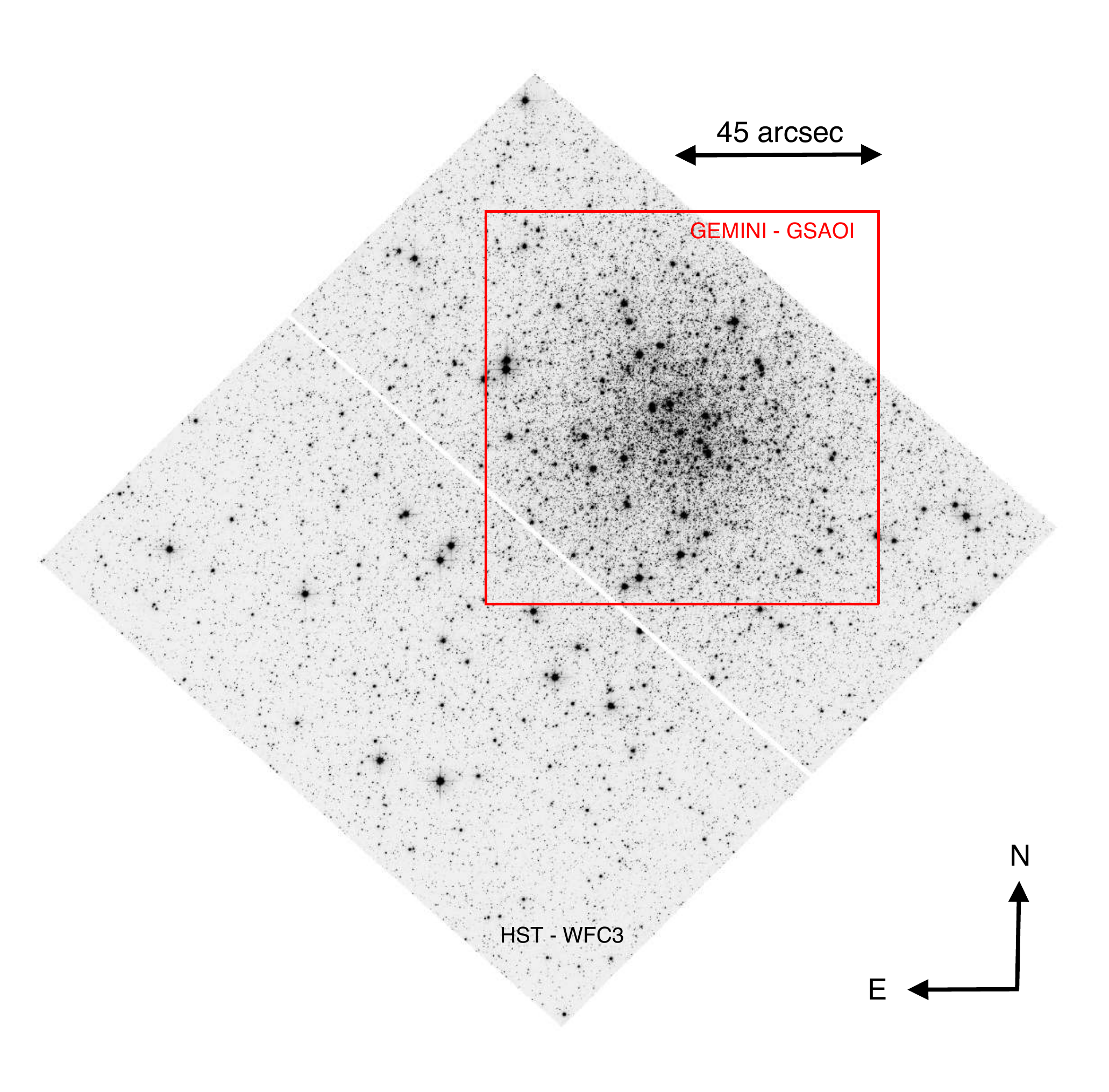}
	\caption{{\it HST}/WFC3 image of NGC 6569 in the $F814W$ filter. North is up, east is left. The WFC3 FOV is 162" x 162". The red box corresponds to the GSAOI/GEMINI FOV, of 92" x 92" on the sky.\label{fig:6569fig1}}
\end{figure}

The optical dataset has been obtained from the UVIS channel of the WFC3, on board {\it HST}. The WFC3 UVIS images are characterized by a spatial resolution of 0.04"/pixels and two twin chips covering a total FOV of 162" x 162" on the sky. The observations have been acquired in 2017, as part of the {\it HST} proposal GO 15232, PI: F.R. Ferraro. They consist of 6 images in $F555W$ with $t_{exp}$ = 30 sec and 6 images in $F814W$ with $t_{exp}$ = 13 sec. Figure \ref{fig:6569fig1} shows a {\it HST}/WFC3 image in the $F814W$ filter, with superimposed the FOV covered by the GEMINI observations discussed above.
\subsection{Data reduction} 
For the NIR dataset, we used standard \texttt{IRAF\footnote{IRAF is distributed by the National Optical Astronomy Observatory, which is operated by the Association of Universities for Research in Astronomy, Inc., under cooperative agreement with the National Science Foundation.}} \citep{tody1986,tody1993} tools to correct the raw images for flat-field and to perform the sky subtraction. A master sky was obtained by combining five sky images of a relatively empty field. For the optical dataset, instead, we used images processed, flat-fielded, bias subtracted and corrected for Charge Transfer Efficiency (CTE) losses by standard {\it HST} pipelines (\_flc images). The WFC3 images have been also corrected by the Pixel Area Map using the files available in the {\it HST} Web site.
For both NIR and optical data, the photometric reduction was carried out via Point Spread Function (PSF) fitting techniques in each chip of each image independently, by using \texttt{DAOPHOTIV} \citep{stetson1987}. The PSF has been modeled by selecting about two hundred bright and isolated stars uniformly distributed in each chip, and by using the \texttt{DAOPHOTIV/PSF} routine. We allowed the PSF to vary within each chip following a cubic polynomial spatial variation. The best-fit PSF analytic models obtained for the {\it HST} $F555W$ and $F814W$ images, are a Moffat function with $\beta$ = 1.5 \citep{moffat1969} and a Penny function \citep{penny1976}, respectively. The typical PSF model for the GEMINI images is a Penny function, but in some cases it was necessary to adopt a Lorentz function \citep{melcher1977} as well. The PSF models thus obtained were then applied to all the star-like sources detected at a 3$\sigma$ level above the local background by using \texttt{ALLSTAR}. We have derived in this way the stellar instrumental magnitudes. Then, starting from the star lists thus obtained and to fill the gaps among the GSAOI chips, we created a GEMINI master star list containing all the stars measured in at least three $K_{s}$ images. For HST instead, the master list was made up of all the stars measured in at least four $F814W$ observations. As done in previous works (e.g. \citealp{dalessandro2014} and references therein), the master lists thus created for the two datasets have been used as input for \texttt{ALLFRAME} \citep{stetson1994}. The files obtained as output have been independently combined to get a GEMINI catalog containing all the stars measured in at least three $J$ and three $K_{s}$ images and an {\it HST} catalog containing stars measured in at least four $F555W$ and four $F814W$ exposures. For every stellar source in each catalog, different magnitude estimates have been homogenized and their mean values and standard deviations have been adopted as the star magnitudes and photometric errors in the final catalog (\citealp{ferraro1991,ferraro1992}). %This has been done for NIR and optical catalogs separately.
\subsection{Calibration and astrometry}  
The $J$ and $K_{s}$ instrumental magnitudes have been converted into the 2MASS photometric system by using the stars in common with the VVV catalog of NGC 6569 \citep{mauro2012,cohen2017} as secondary photometric calibrators. The $F555W$ and $F814W$ instrumental magnitudes have been instead reported to the VEGAMAG photometric system by adopting the zero points listed in the WFC3 web site\footnote{http://www.stsci.edu/hst/wfc3/phot\_zp\_lbn} for a 0.4" (10 pixels) aperture correction. Both catalogs have been finally roto-translated to the absolute (RA, Dec) coordinates using the stars in common with the VVV catalog\footnote{The transformation into the absolute reference frame has been double-checked using the stars in common with the Gaia Data Release 2 catalog.} and the cross-correlation software \texttt{CataXcorr}.
\subsection{GEMINI performances}
\begin{figure}[ht]
	\figurenum{3}
	\plotone{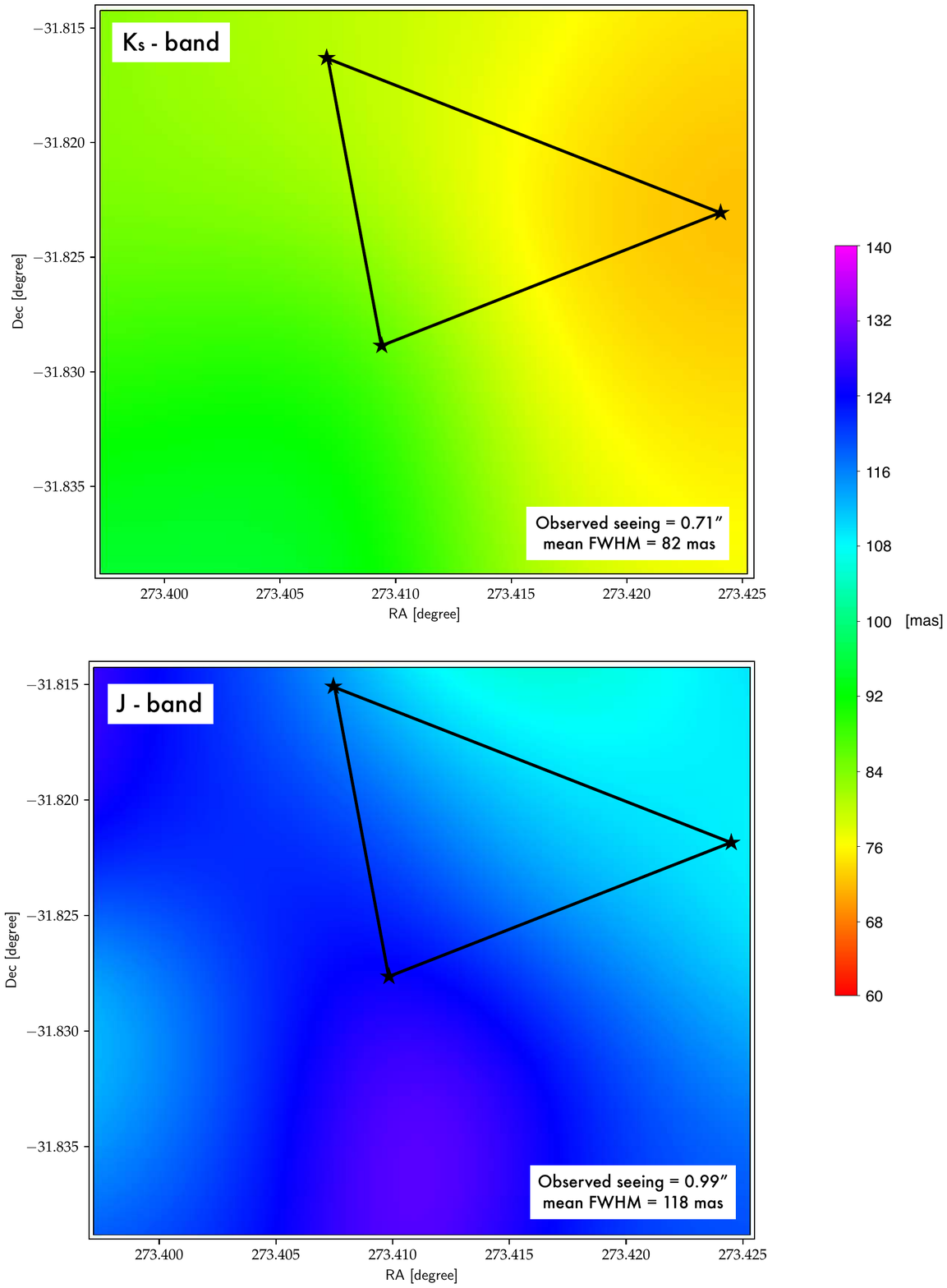}
	\caption{FWHM maps of the best $K_{s}$ image (top panel) and the worst $J$ image (bottom panel). They are based on a sample of $\sim$ 450 and 300 stars, in the corresponding mosaic, co-added images. Average FWHM values of 82 mas and 118 mas have been derived for the best and the worst maps, respectively. The tip-tilt NGS stars are presented by black star symbols and connected through solid lines. The observed seeing, after correction for airmass, is marked in the maps. The color code, ranging from magenta (worst) to red (best), is a performance indicator. As can be seen, the best correction is achieved closer to the NGS asterism.\label{fig:6569fig2}}
\end{figure}

By following the approach used in \citet{dalessandro2016}, we have analyzed the quality of the GEMINI data of NGC 6569. The average Full Width at Half Maximum (FWHM) varies from 3.5 to 5.0 pixels (70 mas - 100 mas) for the images acquired in the $K_{s}$ band and from 4.0 to 7.0 pixels (80 mas - 140 mas) in $J$. For reference, the diffraction limited FWHMs are 68 mas and 37 mas at 2.2 $\mu m$ ($K_{s}$-band) and 1.2 $\mu m$ ($J$-band), respectively. We have verified that the FWHM is quite stable across the FOV with a maximum variation of 10-15\% in $K_{s}$ and up to 20\% in $J$, in good agreement with what is found in previous works using GEMINI (\citealp{saracino2015,saracino2016}, \citealp{massari2016}). We also verified that the quality of our images strongly depends on the seeing at the sky position of the target (i.e. at the observed airmass), as already shown in Figure 2 of \citet{dalessandro2016}. We derived the observed seeing as $s(500 nm,z) = 10.31 /(r_0(500nm) \times sec(z)^{-3/5})$ where $r_{0}$ is the Fried parameter and $z$ the observed airmass. In Figure \ref{fig:6569fig2} the best $K_{s}$ and the worst $J$ images (in terms of delivered FWHMs) are presented as an example, with the reference guide stars selected for the tip-tilt correction marked as black star symbols.

\section{NIR and optical CMD\scriptsize{s} \normalsize{of NGC 6569}}
\begin{figure}[ht!]
	\figurenum{4}
	\plotone{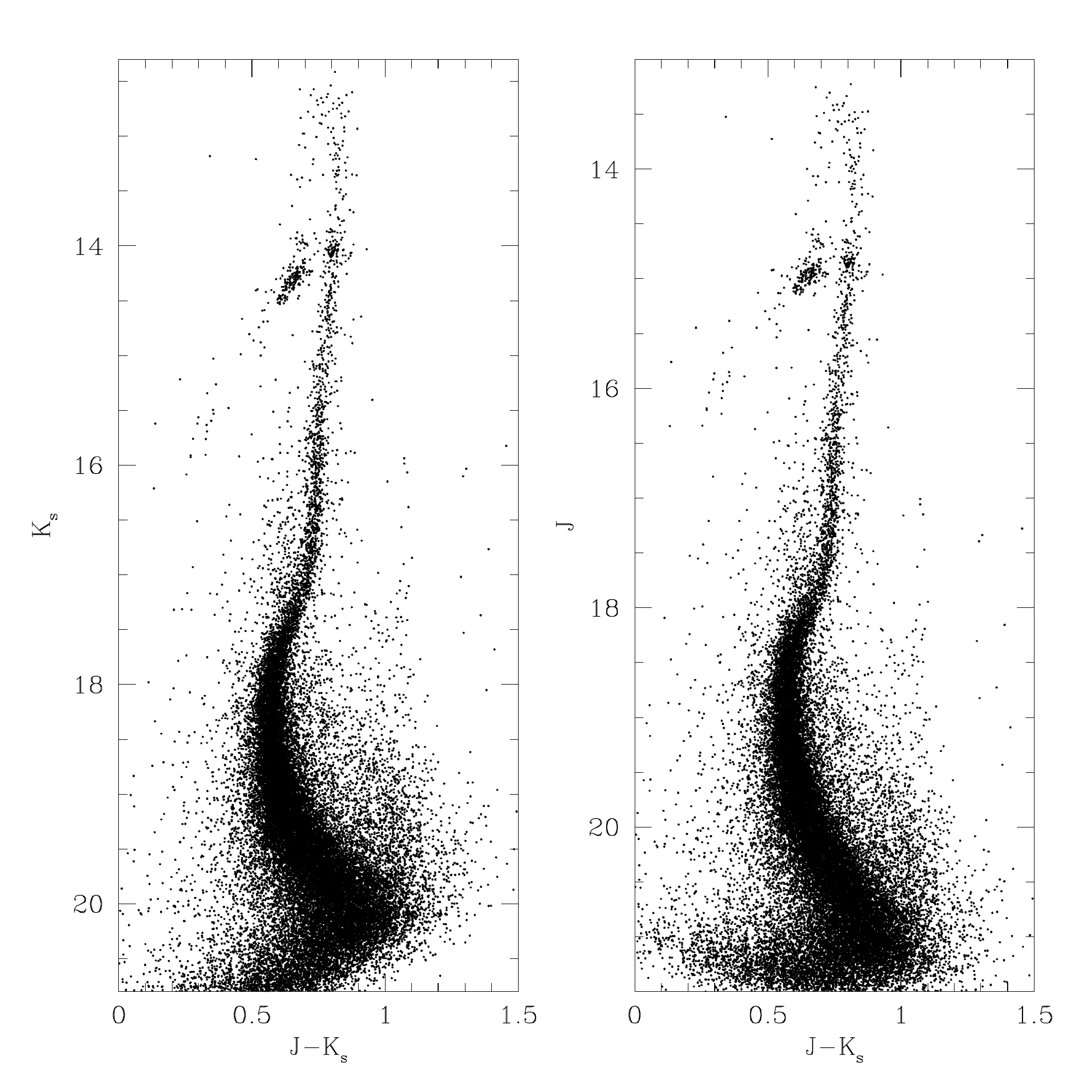}
	\caption{($K_{s}, J-K_{s}$) and ($J, J-K_{s}$) CMDs of NGC 6569 obtained from the GEMINI observations discussed in Section 2. All the main evolutionary sequences of the cluster are clearly visible, from the RGB, HB, down to the MS. The quality and the deepness of these NIR CMDs is comparable to those achieved with {\it HST}/WFC3 observations (see Figure 12 in \citealt{cohen2018}).\label{fig:6569fig3}} %The photometric errors for each bin of $K_{s}$ and $J$ magnitudes are shown on the right side of the panels.
\end{figure}

The ($K_{s}, J-K_{s}$) and ($J, J-K_{s}$) CMDs of NGC 6569 are presented in Figure \ref{fig:6569fig3}. The comparison with Figure 12 of \citet{cohen2018} clearly demonstrates the extraordinary capabilities of the GeMS/GSAOI system, that allowed us to obtain NIR CMDs with quality and deepness comparable to those achieved by {\it HST}/WFC3 observations. It is worth noting that the $J$ band is limiting the overall CMD depth. Nevertheless, the CMD spans $\approx$ 8 magnitudes, ranging from bright red giant branch (RGB) stars down to low-mass MS stars. Unfortunately, the stars lying along the brightest portion of the RGB ($K_{s} <$ 13.5) are saturated in all the available images. Instead, the HB, the SGB and the MS-TO ($K_{s}$ $\approx$ 18.0) are well defined. Unfortunately, our data do not go deep enough to unambiguously identify the MS-Knee \citep{saracino2018}.

\begin{figure}[ht!]
	\figurenum{5}
	\plotone{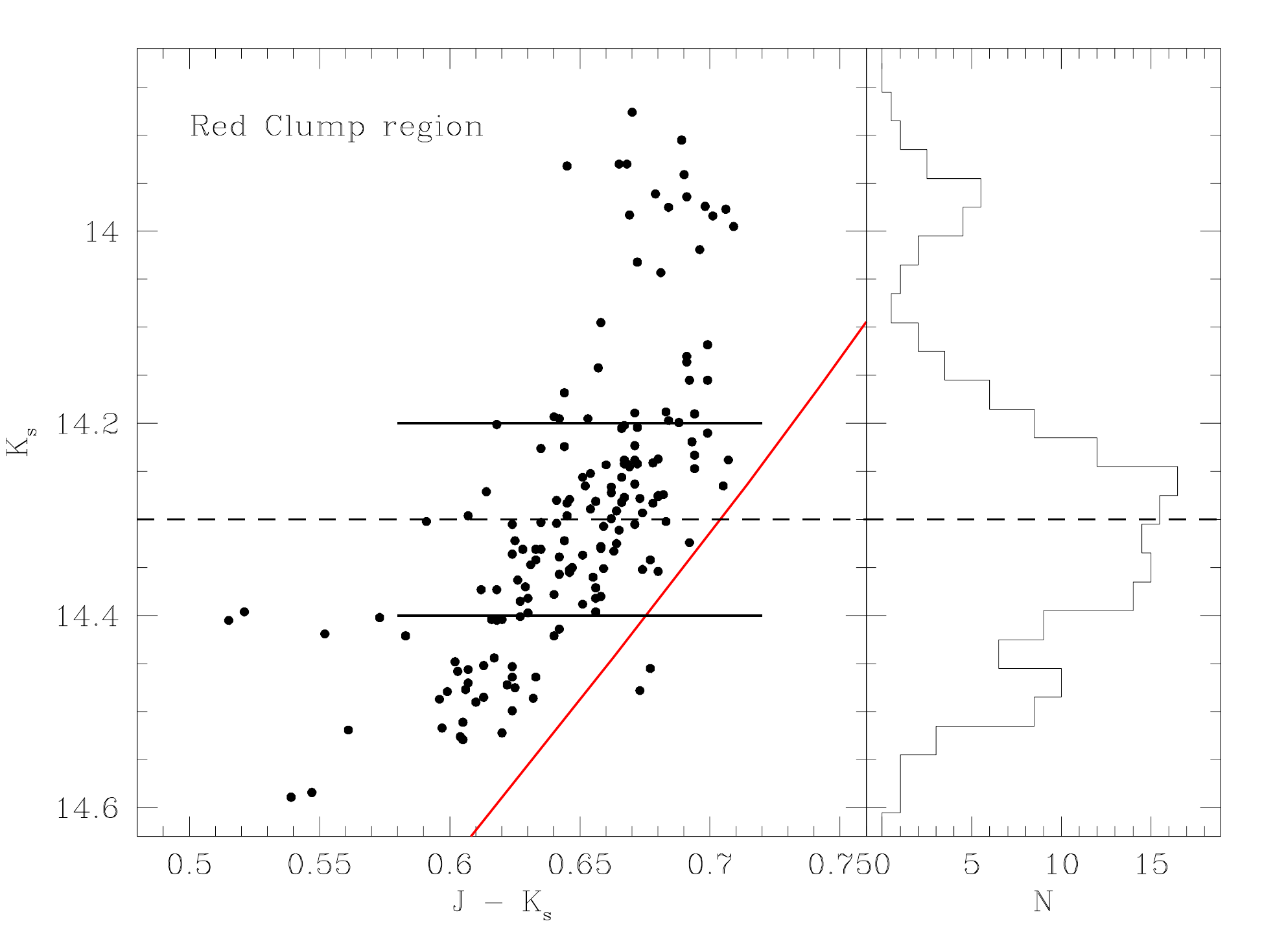}
	\caption{{\it Left panel:} ($K_{s}, J-K_{s}$) CMD zoomed in the red clump region. The two solid lines mark the brightest and the faintest boundaries of the HB selection boxes used by \citet{mauro2012}. The dashed line marks the magnitude level where the red clump shows the gap in the Hess diagram of \citet[][see their Figure 3]{mauro2012}. A Zero Age Horizontal Branch \citep{pietrinferni2004} for the chemical composition of the cluster is shown in the plot as red line just to guide the eye through the HB shape in this filter combination. {\it Right panel:} Magnitude histogram of the stars observed in the red clump region.\label{fig:6569HB}}
\end{figure}

\begin{figure}[ht!]
	\figurenum{6}
	\plotone{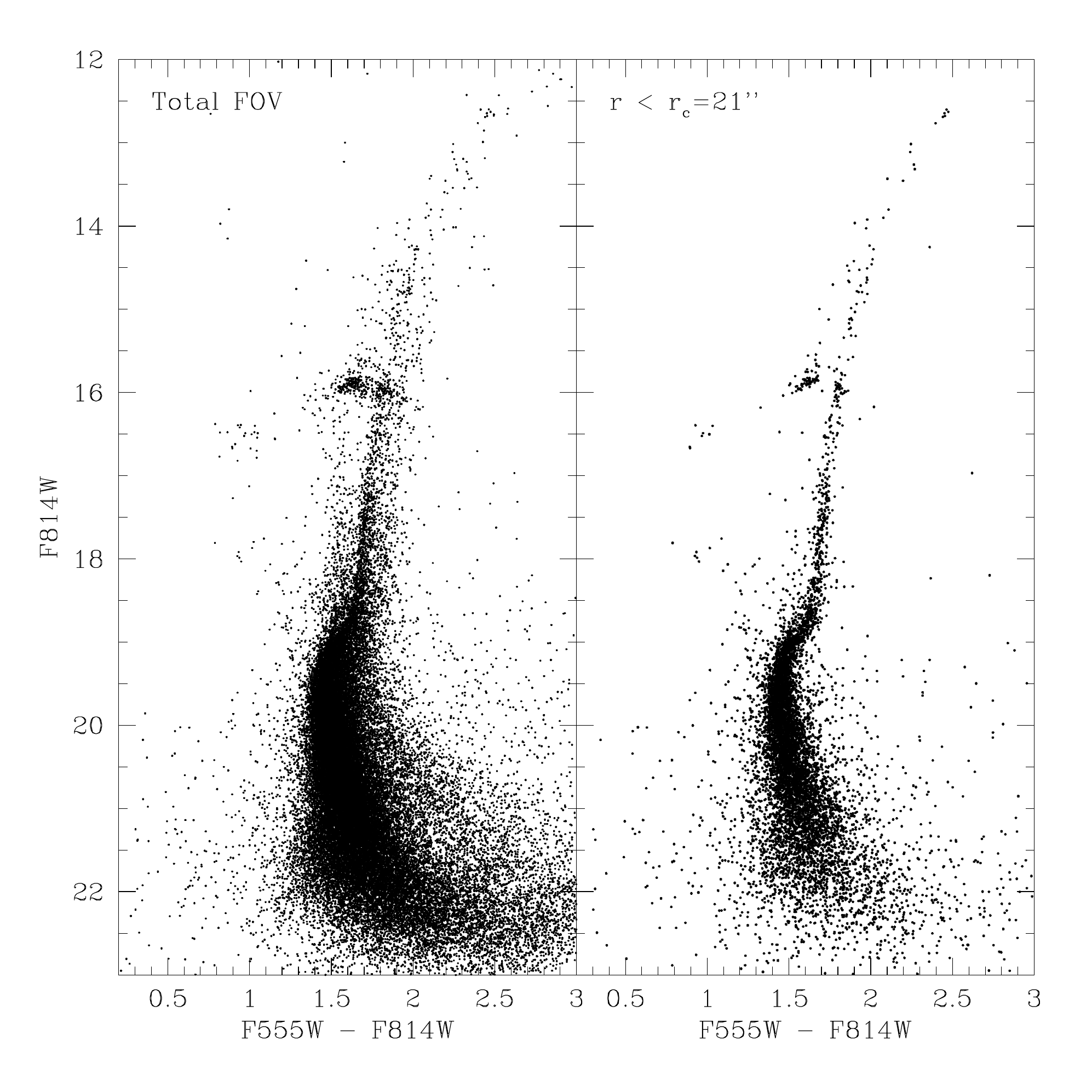}
	\caption{The {\it HST} ($F814W, F555W-F814W$) CMD of NGC 6569 is shown for the total FOV ({\it left panel}) and for r $<$ $r_{c}$ ({\it right panel}), in order to amplify the differences observed in the central region of the cluster, compared to a more external one. As can be seen, the optical CMD extends for more than 10 magnitudes but field contamination starts to be important going outwards.\label{fig:6569fig4}}
\end{figure}

The ($K_{s}, J-K_{s}$) diagram allows us to investigate the presence of the double red clump claimed by \citet{mauro2012} from the analysis of VVV data. Figure \ref{fig:6569HB} shows a zoom in the red clump region, with the horizontal lines marking the magnitude limits used by \citet{mauro2012} to separate what they named HB-A and HB-B (see their Figure 3). As apparent, our photometry shows no strong evidence of a double HB in this cluster. However, a one-to-one comparison is not really possible because the VVV photometry of \citet{mauro2012} cover a much larger FOV than the combined GEMINI+{\it HST} data but it is severely affected by crowding in the core. Instead, in agreement with \citet{cohen2018}, we confirm the presence of a blue extension of the HB, that is somehow unexpected for a quite metal-rich GC as NGC 6569. %The bending observed along the MS at $K_{s}$ = 20.0-20.5 can be identified with the MS-knee (\citealp{bono2010}, \citealp{massari2016}, \citealp{saracino2016,saracino2018}), but photometric errors are too large at these magnitude levels to allow an accurate detection of this feature. 
Sequences of field stars are visible at $J-K_{s}$ $\approx$ 1.0 and running parallel to the RGB at $J-K_{s}$ $\approx$ 0.6.\\
The ($F814W, F555W-F814W$) CMD of the cluster is shown in Figure \ref{fig:6569fig4}. The left panel includes all stars in the WFC3 FOV, while in the right panel we show only stars located at a distance r $<$ $r_{c}$ where $r_{c}$ = 21" is the core radius of the cluster \citep{harris1996}. The optical CMD nicely extends for more than 10 magnitudes, down to $F814W$ $\approx$ 23.0. All the evolutionary sequences are clearly distinguishable and a few blue HB stars at $F814W$ $\approx$ 16.5 can be identified also in this filter combination. Given that the WFC3 has a larger FOV than GSAOI, a larger degree of field contamination is evident and clearly appears also from the comparison of the two panels in Figure \ref{fig:6569fig4}.
\section{Proper Motion Analysis}
To clean the observed CMDs from field star interlopers, we performed a relative PM analysis. We used the GeMS+GSAOI and {\it HST}/WFC3 datasets, which are separated by a temporal baseline of 4.419 years, as first and second epoch, respectively. PM information from the Gaia Data Release 2 for this cluster exist only for $\sim$ 200 stars along the brightest portion of the RGB and suffer from significant uncertainties.\\ 
Our analysis represents a crucial test to verify whether ground-based AO observations can be successfully combined with {\it HST} data to perform accurate PM measurements in dense stellar systems (\citealt{fritz2017} already demonstrated that they work quite well in loose environments as the halo GC Pyxis).

\begin{figure}[ht!]
	\figurenum{7}
	\plotone{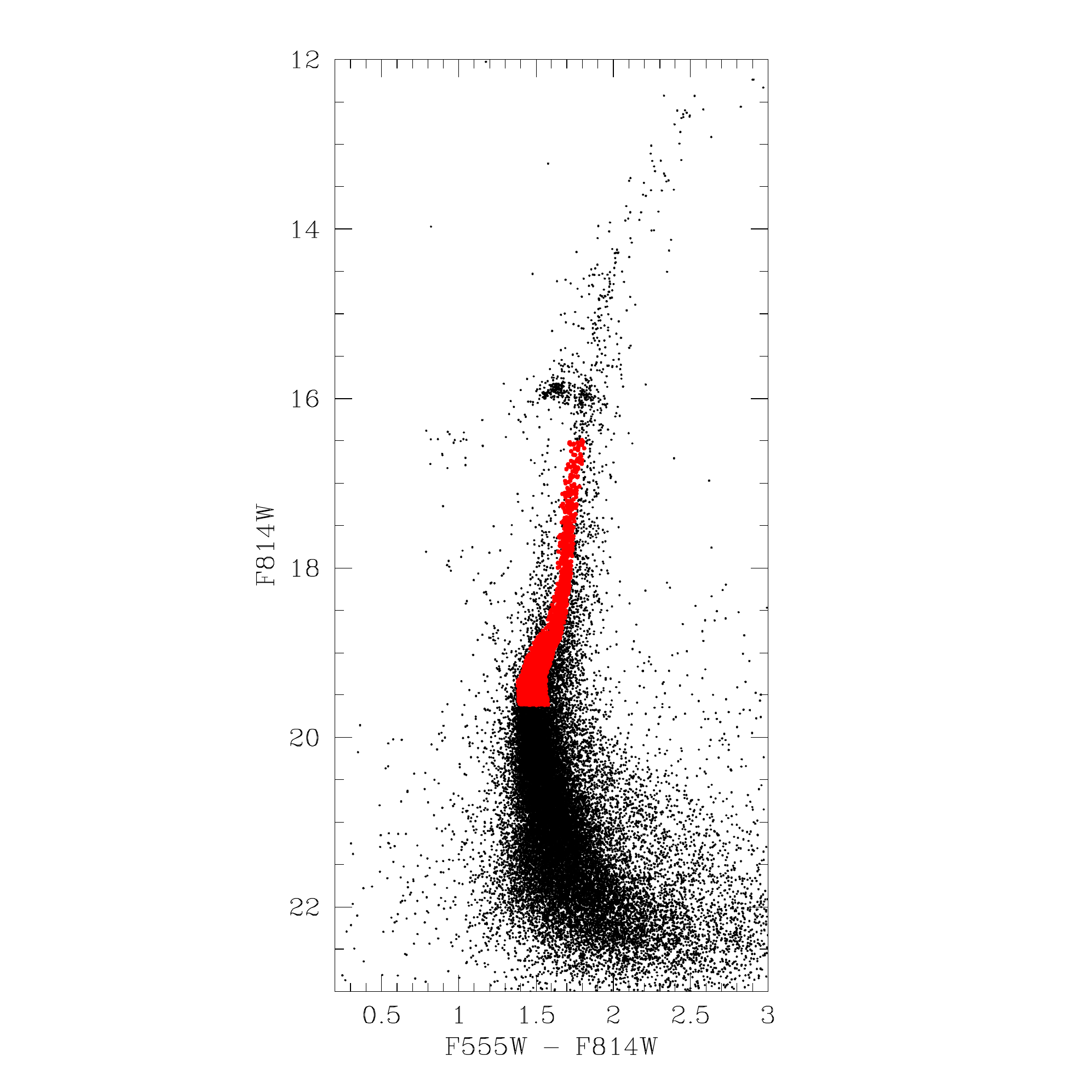}
	\caption{The stars used to create the {\it HST} {\it master frame} are shown in red on top of the optical ($F814W, F555W-F814W$) CMD of NGC 6569. They are likely cluster members on the basis of their position along the lower RGB, the SGB and the upper MS.\label{fig:6569fig5}}
\end{figure}

To measure the stellar PMs we adopted the approach described in \citet[see also \citealp{dalessandro2013}, \citealp{bellini2014}, \citealp{massari2015}, \citealp{cadelano2017}]{massari2013}. Briefly, the procedure consists in measuring the displacement of the centroids of the stars measured in the two epochs, once a common reference frame is defined. The first step is to adopt a distortion-free reference frame, that we call {\it master frame} hereafter. We defined as {\it master frame} the catalog obtained from the combination of all the (second-epoch) WFC3 single-exposure catalogs, after correction for geometric distortions applied following \citet{bellini2011}. Our {\it master frame} contains only stars observed in all the available images (6 for each filter). To derive accurate transformations between the first-epoch and the master catalog, we selected a sample of $\sim$ 3300 bona-fide stars having magnitude 16.5 $\leq$ $m_{F814W}$ $\leq$ 19.5 (corresponding to magnitudes 15.0 $\leq$ $m_{Ks}$ $\leq$ 18.0), which are likely cluster members on the basis of their CMD position (stars distributed along the lower RGB, the SGB and the upper MS), as shown in Figure \ref{fig:6569fig5}. We then applied a six-parameter linear transformation between the two epochs, determined by using the stars in common between the GSAOI and the reference stars in the master catalog.\footnote{For the determination of both PMs and differential reddening (see Section 5) we considered only the FOV in common between the GEMINI observations and the WFC3 chip that contains the cluster center (see Figure \ref{fig:6569fig1}).} We treated each chip separately, in order to maximize the accuracy. Moreover, we carefully corrected the first-epoch GEMINI catalog for the important geometric distortions affecting the GSAOI camera (\citealp{dalessandro2016}, \citealp{massari2016astro}, \citealp{ammons2016}, \citealp{neichel2016}, \citealp{kerber2019}). To do that, we applied the geometric distortions solutions published in \cite{dalessandro2016}.

At the end of this step, for the $\sim$ 3300 selected stars, we have up to 10 position measurements in the first-epoch catalog and up to 12 in the second-epoch catalog. The mean X and Y positions of each star in each epoch is computed by adopting a $\sigma$-clipping rejection. The derived transformations have been then applied to all the stars detected in each frame, by following the same procedure. The relative PM is finally determined by measuring the difference of the mean X and Y positions of the same star in the two epochs, divided by their temporal baseline $\Delta$T = 4.419 years. Such displacements are in units of pixels yr$^{-1}$. %The error on the relative PM of each star is computed by combining the first- and the second-epoch errors as follows: $\sigma^{X}_{PM} = \frac{\sqrt{(\sigma^{X}_{1}+\sigma^{X}_{2})}}{\Delta T}$ and $\sigma^{Y}_{PM} = \frac{\sqrt{(\sigma^{Y}_{1}+\sigma^{Y}_{2})}}{\Delta T}$. 
We finally converted the PMs into absolute units (mas yr$^{-1}$) by multiplying the measured displacements by the pixel scale of the {\it master frame} (0.04"/pixel). On the basis of the measured PMs, we iterated the procedure by refining the {\it master frame} used to compute the six-parameter linear transformation. In particular we selected only stars with PMs ($dx$, $dy$) $<$ 0.3 mas/yr and having magnitude 15.5 $\leq$ $m_{F814W}$ $\leq$ 21.5. At the end we derived relative PMs for $\sim$ 20000 individual stars in the cluster core.

\begin{figure}[ht!]
	\figurenum{8}
	\plotone{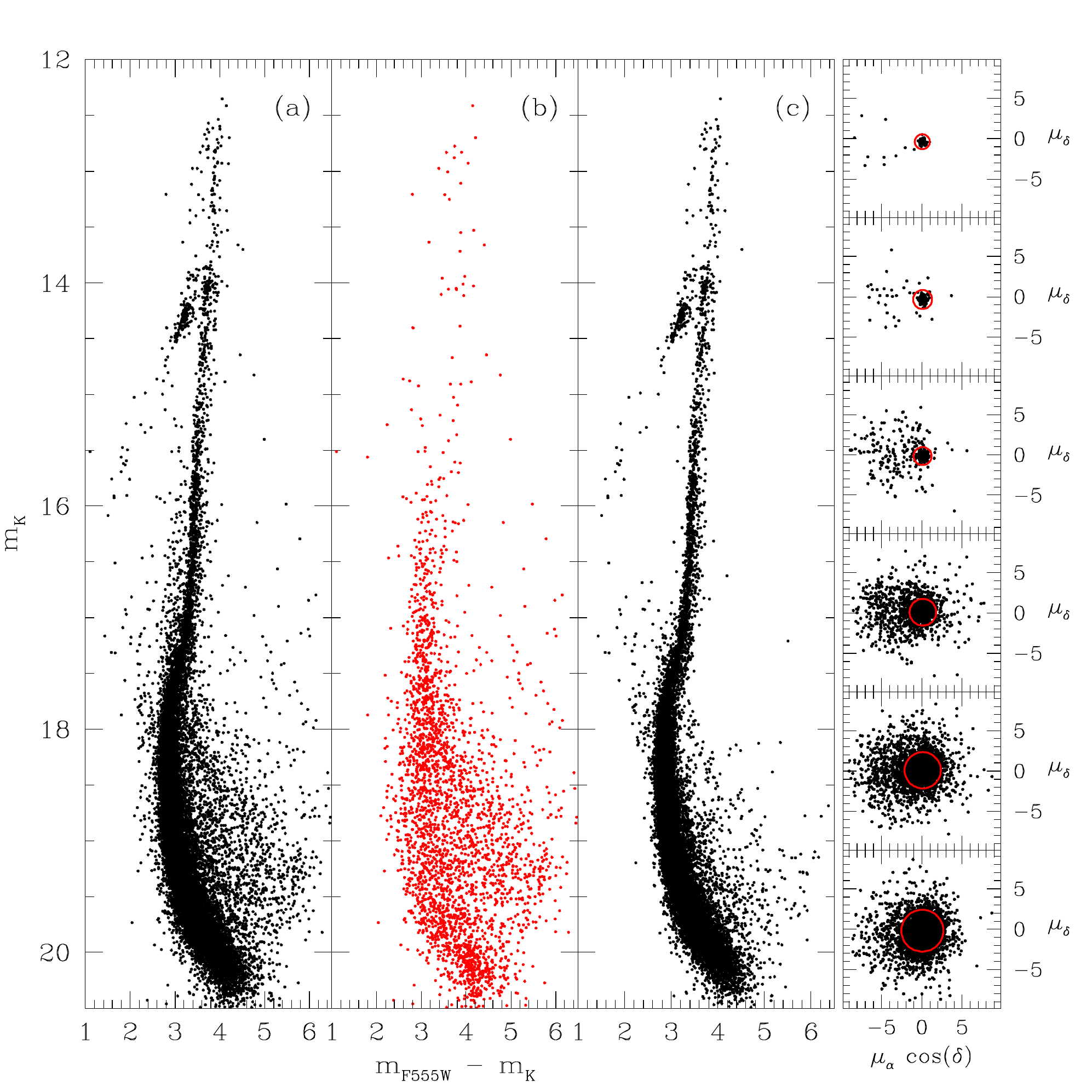}
	\caption{Panel (a): ($m_{K}, m_{F555W}-m_{K}$) CMD of the stars in common between GEMINI and {\it HST}. Panel (b): CMD made of all the contaminating objects, selected from the VPDs as those with PMs not compatible with that of the GC. Panel (c): PM-cleaned CMD obtained by using only the likely cluster members selected from the VPDs shown in the right-most column. As can be seen, sharper and better defined evolutionary sequences are now appreciable. Right-most column: VPDs of the measured stars divided in 6 bins of 1.5 mag each. The solid red circles contain the stars classified as likely cluster members.\label{fig:6569fig7}}
\end{figure}

Figure \ref{fig:6569fig7} shows the derived vector point diagrams (VPDs) and the effect of decontaminating the CMD of NGC 6569 from field stars. The hybrid ($m_{K}, m_{F555W}-m_{K}$) CMD of the stars in common between {\it HST} and GEMINI is presented in the panel (a), before PM selection. The relative PM measurements are shown in the rightmost panels for 6 different bins of 1.5 mag each. In each magnitude interval, we adopted a 2$\sigma$-clipping procedure to separate cluster member from field stars, where $\sigma$ is the dispersion in the VPD. %, defined as the square root of the quadratic sum of the dispersions of the points along the $\mu_{\alpha}cos(\delta)$ and the $\mu_{\delta}$ directions. 
The size of the plotted red circles is proportional to the value of $\sigma$ obtained after convergence. %It increases at fainter magnitudes because both the photometric and the positional errors becomes larger. 
Stars within the red circles are plotted as black points in the panel (c) of Figure \ref{fig:6569fig7}, while the stars located outside the circles are shown in the panel (b).
As apparent in panel (c) of Figure \ref{fig:6569fig7}, all the cluster evolutionary sequences are much better defined after removal of field star interlopers. The contamination observed at the MS level is mostly due to Galactic disk stars, while the narrow sequence of stars observed running parallel on the left of the RGB of NGC 6569 is likely due to bulge field stars. Interestingly, the handful of blue HB stars located at ($F555W-K_{s}$) $\approx$ 1.5 in the hybrid optical-NIR CMD survived the PM-cleaning, thus indicating they are likely cluster members. Also a well defined sequence of Blue Straggler Stars (\citealp{sandage1953}, \citealp{ferraro1992,ferraro1995}) is now clearly distinguishable in the CMD. These stars have been found to be powerful tracers of the dynamical evolution of GCs (see \citealp{ferraro2009b,ferraro2012,ferraro2018}, \citealp{lanzoni2016}): their properties will be discussed in a forthcoming paper.
 
Our approach demonstrates that ground-based AO observations can be successfully used to obtain accurate PMs in dense environments\footnote{The approach presented here is similar to that of \cite{massari2016astro} for NGC 6681 and of \cite{monty18} for NGC 2298 and NGC 3201, but in those cases relative PM measurements were already available from previous works using {\it HST} \citep{massari2013,simpuz16,soto2017}.} (see also \citealt{kerber2019}). %This kind of application is key for the next-generation of ground-based telescopes equipped with MCAO modules and for their use to perform accurate astrometric studies in dense stellar systems.
\\
\section{Differential reddening}
Even after applying the relative PM selection presented in Section 4, the main evolutionary sequences in the CMD are rather broad compared to the photometric error measured at those levels. This is likely due to differential reddening in the sampled FOV. In fact, as already discussed in Section 1, NGC 6569 has a quite large color excess $E(B-V)$ $\approx$ 0.53 \citep{harris1996} and because of its location in the Galactic bulge, it can be affected by significant differential reddening even on the small scales of our observations. 
We determined the differential reddening correction for all the optical ($F555W$ and $F814W$) and NIR (J and $K_{s}$) filters, by using only the stars that survived the PM selection. We focused on the hybrid ($m_{K}, m_{F555W}-m_{K}$) CMD for two main reasons: {\it 1.)} the difference $m_{F555W}-m_{K}$ is the largest color baseline we can sample with the available filters (this is important to amplify the differential reddening), and {\it 2.)} these effects are significantly smaller in $K_{s}$ than in $F555W$ (thus, at a first approximation, they mostly produce color shifts in the hybrid plane). 

\begin{figure}[ht!]
	\figurenum{9}
	\plotone{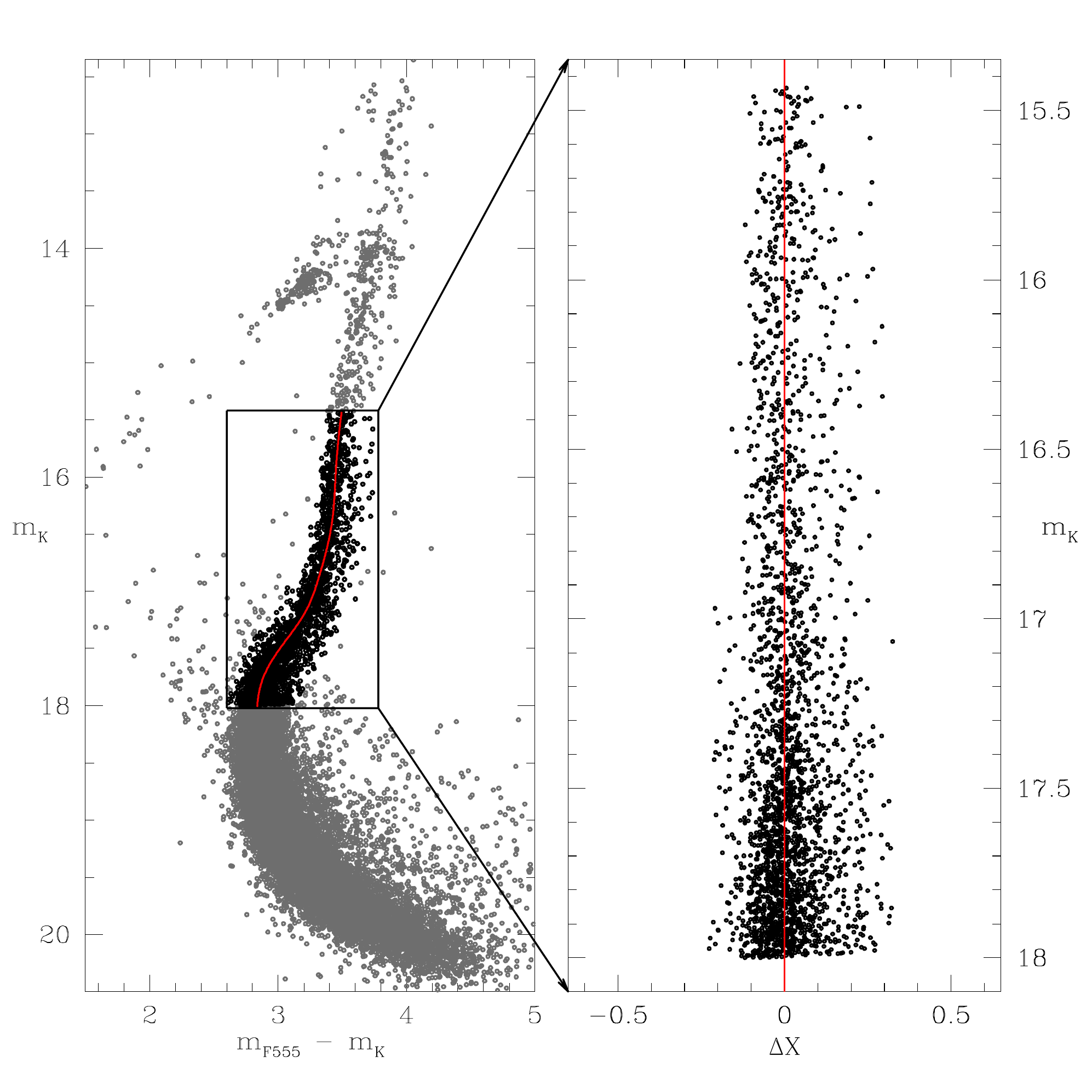}
	\caption{{\it Left panel}: The PM-corrected NIR-optical CMD of NGC 6569. The fiducial cluster MRL for 15.5 $\leq$ $m_{K}$ $\leq$ 18.0 is drawn in red and all the stars used for the differential reddening calculation are encircled by the black box and shown as black dots. {\it Right panel:} Reference stars used for the differential reddening determination (those shown in black in the left panel) plotted in a rectified plane, where the x axis represents the color difference of each star from the MRL, which has $\Delta X$ = 0 by definition.\label{fig:6569fig9}}
\end{figure}

\begin{figure}[ht!]
	\figurenum{10}
	\plotone{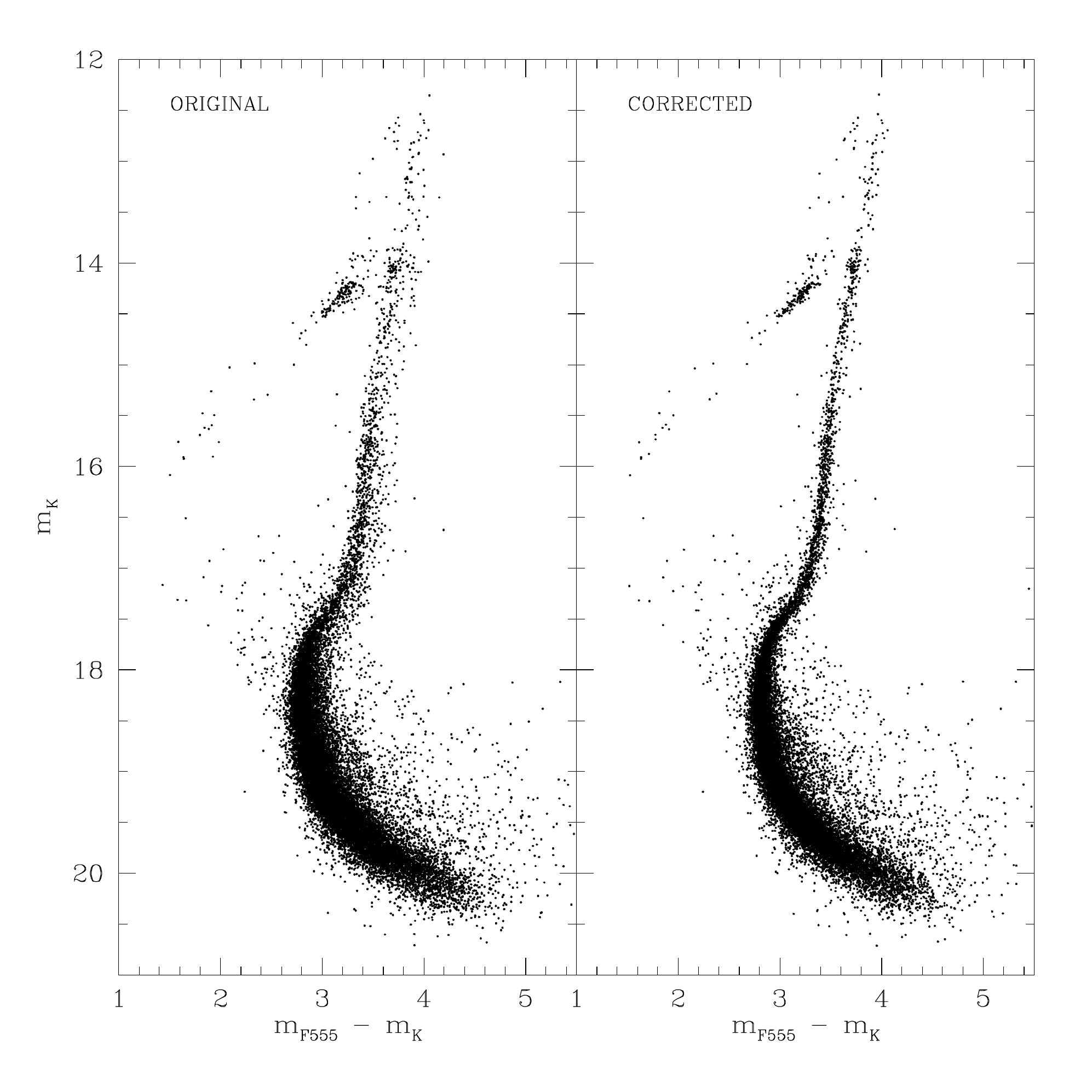}
	\caption{The PM-cleaned ($m_{K}, m_{F555W}-m_{K}$) CMD of NGC 6569 for the stars in common between GEMINI and {\it HST} is shown before (left panel) and after (right panel) the differential reddening correction. In the differential reddening-corrected CMD all the evolutionary sequences are better defined.\label{fig:6569fig10}}
\end{figure}

\begin{figure}[ht!]
	\figurenum{11}
	\plotone{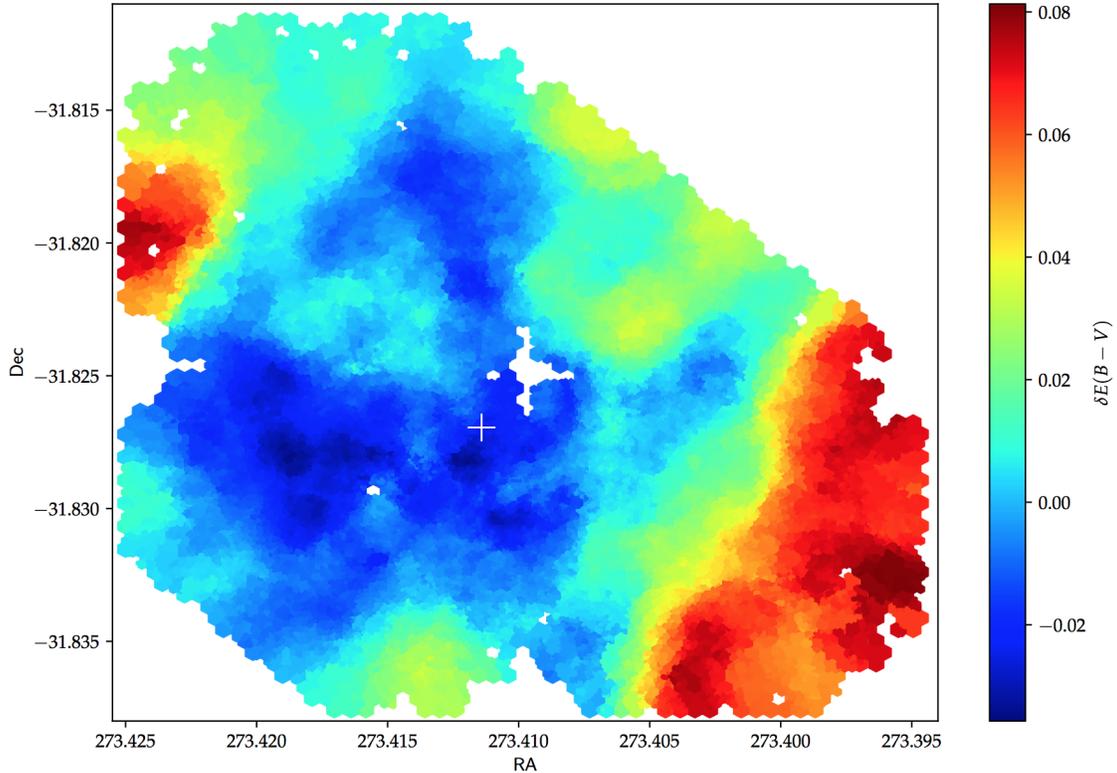}
	\caption{Differential reddening map of NGC 6569 in the GEMINI+{\it HST} FOV. The FOV has been binned in cells of $\sim$ 2 square arcsec, and the $\delta E(B-V)$ value of each cell has been obtained by averaging the color excesses of the stars included. Different colors corresponds to different values of $\delta E(B-V)$, as coded in the adjacent color bar. A clumpy structure is apparent. White cells correspond to saturated stars and/or uncovered regions, while the white cross marks the position of the cluster center.\label{fig:6569fig11}}
\end{figure}

In order to estimate the differential reddening we first derived the cluster mean ridge-line (MRL) in the magnitude range 15.5 $\leq$ $m_{K}$ $\leq$ 18.0 (see Figure \ref{fig:6569fig9}, left panel). To this end, we divided the considered portion of the CMD in different magnitude bins, and in each of them we computed the mean color after a 2$\sigma$-clipping rejection. We adopted five different bin widths, ranging from 0.2 to 0.4 mag by step of 0.05 mag. In each of these five realizations a MRL was obtained. The final MRL is then given by the average of the five values after a re-sampling with a cubic spline of 0.01 mag steps (red line in Figure \ref{fig:6569fig9}, left panel). For all the stars used to derive the MRL (black dots in Figure \ref{fig:6569fig9}, left panel) we then computed the geometrical distance ($\Delta$X) to the MRL. The distribution of $\Delta$X as a function of magnitude for the reference sample is shown in the right panel of Figure \ref{fig:6569fig9}. This reference sample has finally been used to assign a $\Delta$X value to all the stars in our photometric catalog: for every source, $\Delta$X has been determined as the average, with a 2$\sigma$-clipping rejection, of the $\Delta$X values measured for the 20 spatially-closest reference stars\footnote{This number can be considered as the best compromise between having a good statistics and a good spatial resolution. The result does not change significantly if a slightly different number of nearest reference stars (from 10 to 30) is adopted.}. The observed $\Delta$X can then be easily transformed into the local differential reddening, $\delta E(B-V)$, by using the following equation:
\begin{equation}
\delta E(B-V) = \frac{\Delta X}{R_{F555W}-R_{K}}
\end{equation}
where the adopted extinction coefficients are $R_{F555W}$ = 3.274 and $R_{K}$ = 0.366 (see Table A1 in \citealp{cv2014}).
The $\delta E(B-V)$ values were then applied to all the available filters. We show in Figure \ref{fig:6569fig10} the ($m_{K}, m_{F555W}-m_{K}$) CMD of NGC 6569 before (left), and after the differential reddening correction (right).
The differential reddening map of NGC 6569 in the overlapping GEMINI and {\it HST} FOVs is shown in Figure \ref{fig:6569fig11}. In this region $\delta E(B-V)$ varies by $\sim$ 0.12 mag, from -0.036 up to 0.081, with a clumpy structure.
The right panel of Figure \ref{fig:6569fig10} shows the nice CMD obtained after the differential reddening correction (and PM cleaning). All the main evolutionary sequences are now much better defined. We confirm that, after these corrections, the cluster unambiguously shows an HB morphology with a single red clump and a quite extended tail toward blue colors.

\section{Distance determinations to NGC 6569}
An estimate of the distance modulus of NGC 6569 can be derived from the RR Lyrae stars that belong to the cluster. To this purpose, we used the mean $V$ magnitude ($<V>$ = 17.36) of the 25 RR Lyrae presented in the OGLE catalog and the very recent $M_{V}$ - [Fe/H] relations calibrated for this type of stars from the \citet{gaia2017} based on Tycho-Gaia Astrometric Solution (TGAS). Since three different methods have been used to perform such a calibration, we adopted the mean $<M_{V}>$ solution and three main sources of uncertainties: {\it 1)} $\sim$ 0.10 dex in [Fe/H], {\it 2)} the stochastic effects in the fits (0.04 mag), {\it 3)} and the systematics due to the different solutions (0.10 mag). With the assumed metallicity [Fe/H] = $-0.87$, we obtain $<M_{V}>$ = $0.72 \pm 0.11$ for the considered 25 RR Lyrae stars and thus a distance modulus in the $V$ band $(m-M)_V = 16.64 \pm 0.11$. By assuming $R_{V} = 3.1$ as extinction coefficient in the V band, the true distance modulus of NGC 6569 becomes $(m-M)_0 = 15.00 \pm 0.11$. This value  varies by only $\pm$ 0.01 mag when the sub-sample of 7 RR Lyrae stars in common between our PM-cleaned {\it HST} dataset and the OGLE catalog \citep{soszynski2017} is considered. 

\begin{figure}[ht!]
	\figurenum{12}
	\plotone{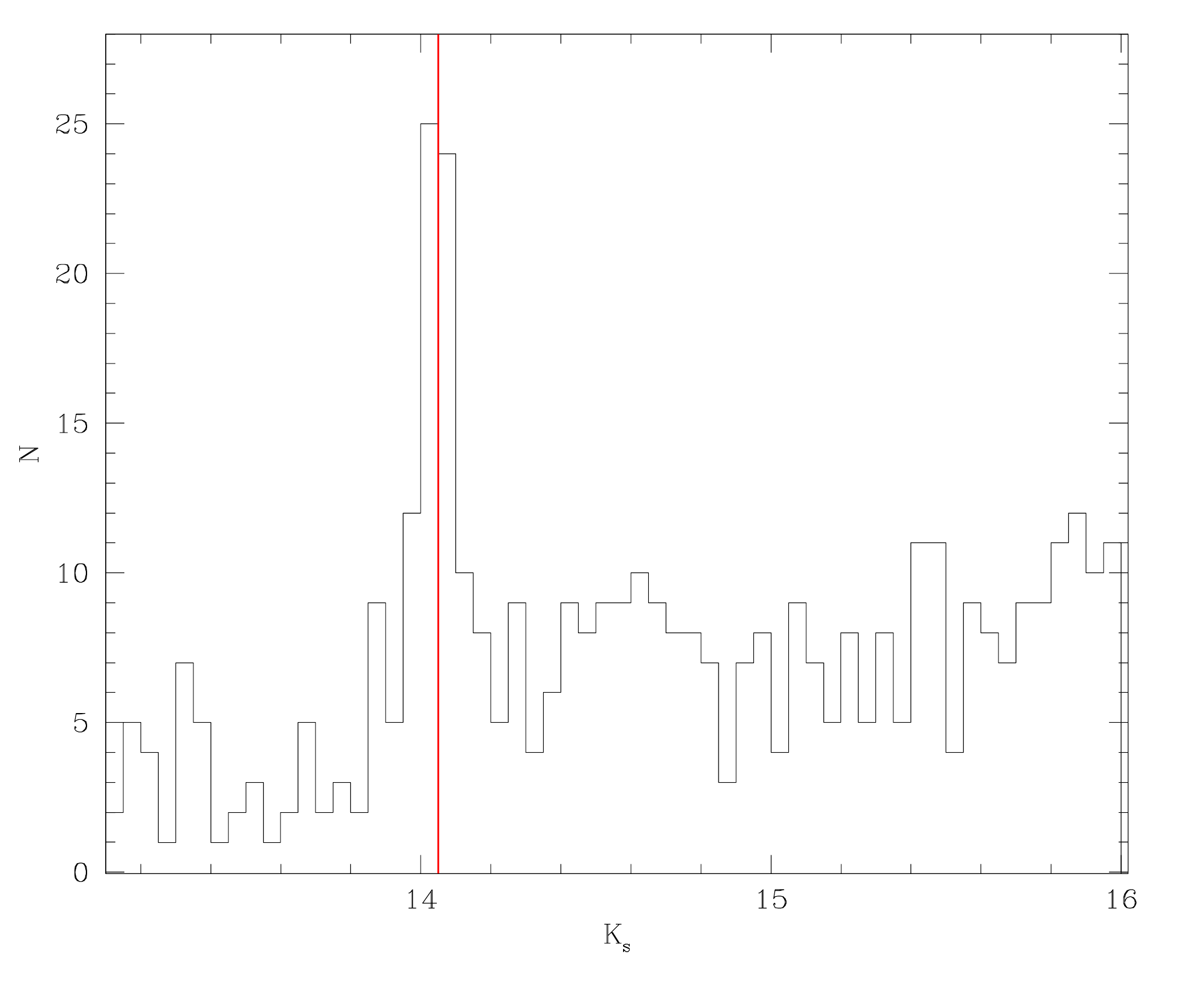}
	\caption{Luminosity function of the bright RGB stars in NGC 6569. The red vertical line marks the location of the RGB-bump.\label{fig:6569figg11}}
\end{figure}

The right panel of Figure \ref{fig:6569fig10} also shows a well defined clump of stars slightly brighter than the red clump along the now narrow RGB: this is the so-called RGB-bump, an evolutionary feature that flags the moment when the H-burning shell reached the H discontinuity left from the inner penetration of the convective envelope (see \citealp{fusipecci1990}, \citealp{ferraro1999,ferraro2000}). Figure \ref{fig:6569figg11} shows the differential luminosity function of the bright RGB stars in the $K_{s}$ band, where the RGB-bump can be easily detected at $K_{s} = 14.05 \pm 0.05$ mag. This feature was identified in the NIR CMDs of 24 Galactic GCs by \citet{valenti2004}, who provided a relation linking the absolute $K_{s}$ magnitude of the RGB-bump and the global cluster metallicity $[M/H]$: $M_{K} = -0.17 + 2.07 [M/H] +0.49 [M/H]^2$.
Since the metallicity of NGC 6569 is well constrained both in terms of iron and $\alpha$-elements, this relation can be used for an independent estimate of the cluster distance modulus.
By adopting [Fe/H] = $-0.87$ and [$\alpha$/Fe] = $+0.4$ and the relation between the iron abundance and the global metallicity presented in \citet{salaris1993}, we obtain [M/H] = $-0.58$. From the \citet{valenti2004} relation we thus obtain $M_{K} = -1.21 \pm 0.12$, from which the distance modulus in the $K_{s}$ band can be derived: $(m-M)_K = 15.26 \pm 0.13$. By assuming an average reddening of $E(B-V)$ = 0.53 \citep{ortolani2001} in the direction of the cluster and the extinction coefficients adopted in Section 5, the true distance modulus of NGC 6569 turns out to be $(m-M)_0 = 15.07 \pm 0.13$, in excellent agreement with the previous estimate.

The position in magnitude of the RGB-bump depends both on the metallicity and the cluster age, but the finding of consistent distance values using independent methods is a proof that the age of this cluster is similar (i.e. old) to that of the GCs used by \citet{valenti2004} for the calibration of the $M_{K}$ - [M/H] relation. 
We thus adopt the weighted mean of the two derived values as the distance modulus of NGC 6569: $(m-M)_0 = 15.03 \pm 0.08$, corresponding to $10.1 \pm 0.2$ kpc from the Sun. This is also in very good agreement with the previous determination of \citet{ortolani2001}, who quote $(m-M)_0 = 14.96 \pm 0.20$, but smaller than the distance of 10.9 kpc given by \citet{harris1996}.

\section{Age estimate}
In the following analysis we use the PM-cleaned and differential reddening-corrected ($m_{F814W}, m_{F555W}-m_{F814}$) CMD of NGC 6569 to derive the cluster absolute age via the isochrone fitting method. The adopted approach allows us to simultaneously estimate the age, distance modulus and color excess of the cluster through a one-to-one comparison between the observed CMD and a set of theoretical models of suitable chemical composition, exploring reasonable grids of values for each of the three parameters (see also \citealp{kerber2018,kerber2019}, \citealp{correnti2016,correnti2018}). % The final aim of the approach is to accurately derive the absolute age of the cluster itself.

The observed CMD has been compared to three different sets of $\alpha$-enhanced isochrones: the Dartmouth Stellar Evolutionary Database (DSED, \citealp{dotter2008}), the Victoria-Regina Isochrone Database (VR, \citealp{vandenberg2014}) and a Bag of Stellar Tracks and Isochrones (BaSTI, \citealp{pietrinferni2004}).
For each family of models we assumed [Fe/H] = $-0.87$, $[\alpha/Fe]$ = $+0.4$, and a helium mass fraction Y=0.2522 (0.253 for the BaSTI isochrones). Since no age estimates are available in the literature for NGC 6569, we explored a wide range of values appropriate for an old system, uniformly distributed from 10 to 15 Gyrs, in steps of 0.5 Gyr. We then allowed the color excess $E(B-V)$ to vary by $\pm 0.1$ in steps of 0.02 mag from the previously adopted value (0.53, \citealp{ortolani2001}), assuming a uniform prior probability distribution. Finally, we allowed the distance modulus to vary by $\pm$ 0.2 mag following a normal distribution with peak value and dispersion equal to the values obtained in the previous section (namely, 15.03 and 0.08, respectively).  

From an operational point of view, we determined the MRL in the optical CMD, by measuring the mean color (after a 2$\sigma$-clipping rejection) in bins of 0.2 magnitudes along the main cluster evolutionary sequences, in the range 15.5 $\leq$ $m_{F814}$ $\leq$ 21.0 where both saturation effects and photometric errors are small. In the same magnitude range, each isochrone has been re-sampled in steps of 0.2 mags using a cubic spline.    
 
To determine the best-fit parameters, as well as to study the confidence intervals and the correlations among them, we applied the Markov Chain Monte Carlo (MCMC) sampling technique. For that purpose, we used the \texttt{emcee} \citep{foremanmackey2013} code to sample the posterior probability in the three-dimensional parameter space, assuming the likelihood $\propto \exp(-\chi^{2}/2)$.%, with the uniform prior probability within the allowed physical ranges for the age and $E(B-V)$, and a normal prior probability in the case of $(m-M)_{0}$. 

The $\chi^{2}$ value is computed as follows:
\begin{equation}
\chi^{2} = \sum_{i=1}^{N_{\text{bin}}} \Biggl[\frac{(\text{color})_{\text{obs},i}-(\text{color})_{\text{mod},i}}{\sigma_{\text{color}, \text{obs}}}\Biggr]^{2} %+ a + b
\end{equation} 
where $N_{\text{bin}}$ indicates the number of magnitude bins along the MRL. $(\text{color})_{\text{obs},i}$ and $(\text{color})_{\text{mod},i}$ are, respectively, the observed and the theoretical ($F555W-F814W$) mean colors in the $ith$ magnitude bin along the MRL, and $\sigma_{\text{color}, \text{obs}}$ is the standard deviation associated with the observed mean color.
%Terms $a$ and $b$ have been added to the $\chi^{2}$ calculation, in order to weigh for two additional observational constraints: the magnitude levels of the RGB bump and the HB, respectively.
%\begin{equation}
%a = \Biggl[\frac{mag_{bump}^{obs}-mag_{bump}^{mod}}{0.1}\Biggr]^{2}
%\end{equation}
%where $mag_{bump}^{obs}$ is the observed magnitude of the RGB bump, while $mag_{bump}^{mod}$ indicates the theoretical prediction for this feature. In the optical CMD the RGB bump extends up to 0.2 mag, with a mean value of $m_{F814W}$ = 15.95 mag. 
%This term is important when the difference in magnitude between the observed and the predicted bump is smaller than 0.1 mag, becoming negligible elsewhere. The term $b$ is defined as:
%\begin{equation}
%b = \Biggl[\frac{mag_{HB}^{obs}-mag_{HB}^{mod}}{0.05}\Biggr]^{2}
%\end{equation}
%where $mag_{HB}^{obs}$ and $mag_{HB}^{mod}$ represent the observed and the predicted magnitudes of the HB, respectively. To determine the predicted level of the HB, we adopted as a reference the Zero Age Horizontal Branch (ZAHB) models of \citet{pietrinferni2004} for [Fe/H] = -0.87. Adding this term, we allowed the HB magnitude to vary within a range of $\pm$ 0.05 mag from its observed value ($m_{F814W}$ = 15.90 mag)

\begin{figure}[!ht]
	\figurenum{13}
	\plotone{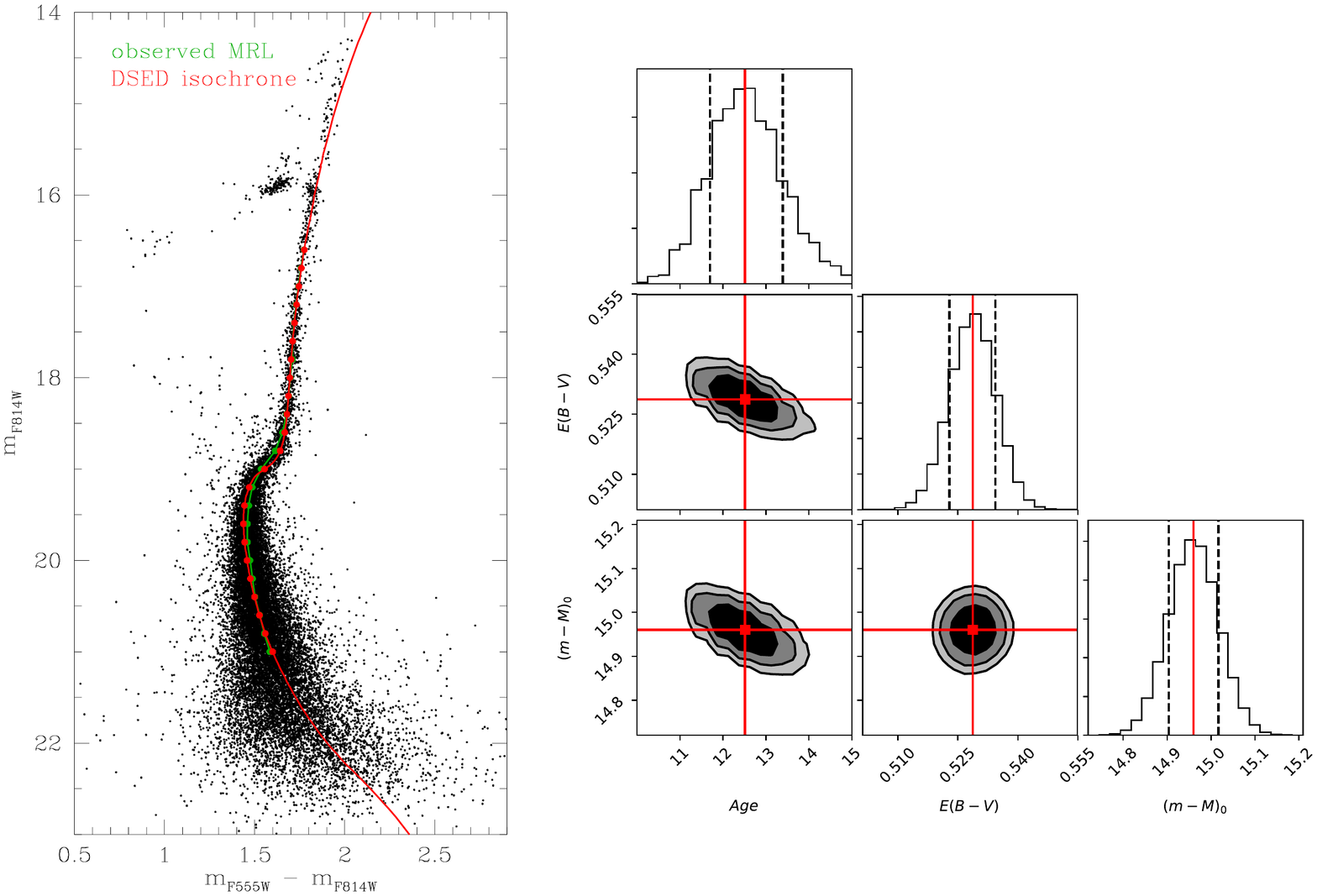}
	\caption{{\it Left panel}: The fiducial line of NGC 6569 is superimposed in dark green on the optical CMD of NGC 6569. In addition, the best-fit solution for the DSED set of isochrones is shown as a red line. The best-fit parameter values are presented in Table \ref{6569_tab1}, left column. The green and red points mark, respectively, the observed and the theoretical color values used to measure the $\chi^{2}$. {\it Right panel}: Corner plots showing the one- and two-dimensional projections of the posterior probability distributions for all the parameters derived from the MCMC method for the DSED isochrones. The contours correspond to the 1$\sigma$, 2$\sigma$ and 3$\sigma$ levels. \label{fig:6569fig12}}
\end{figure}

\begin{figure}[!ht]
	\figurenum{14}
	\plotone{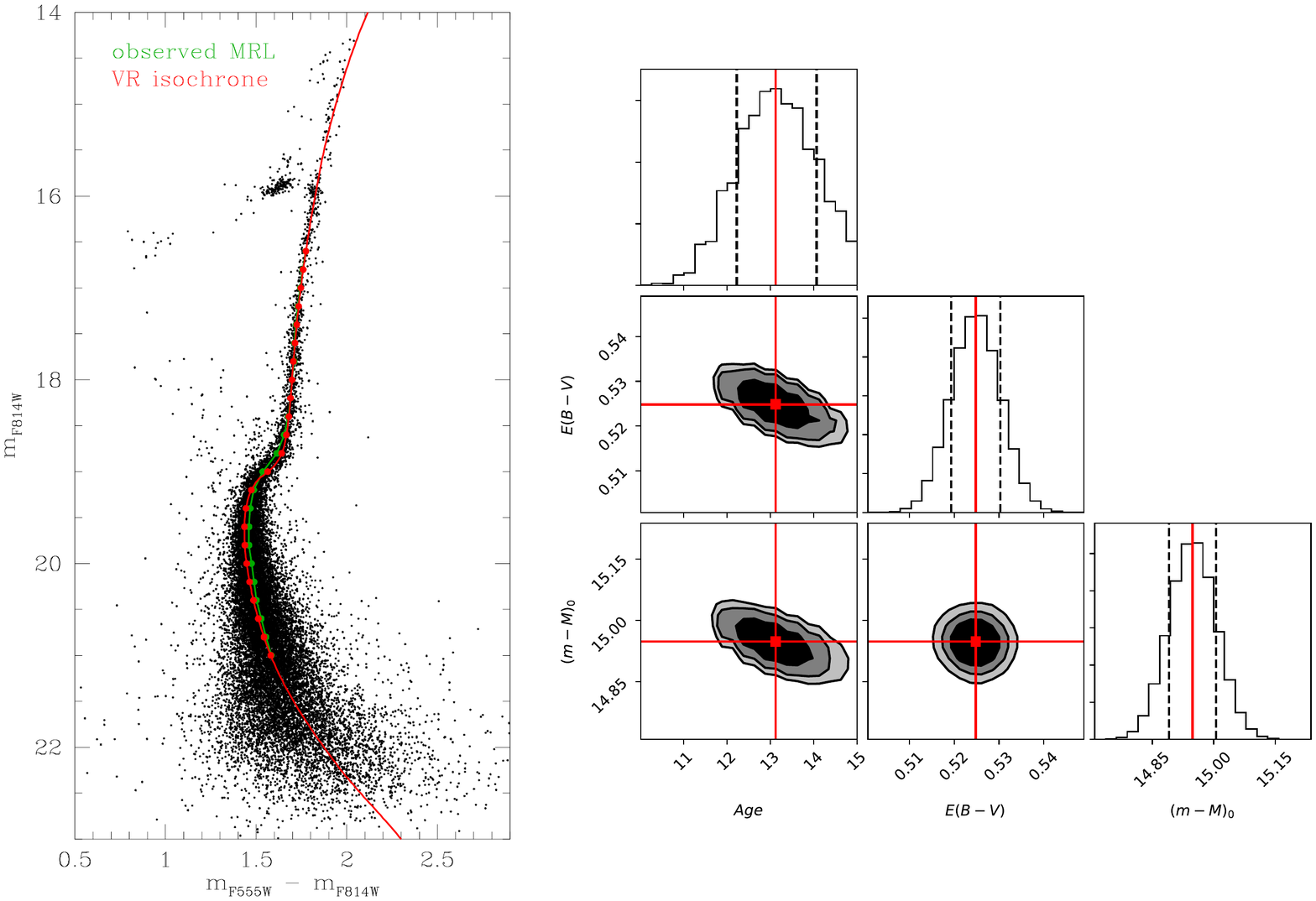}
	\caption{Similar to Figure \ref{fig:6569fig12} but for the VR models. The best-fit parameter values are presented in Table \ref{6569_tab1}, middle column.\label{fig:6569fig13}}
\end{figure}

\begin{figure}[!ht]
	\figurenum{15}
	\plotone{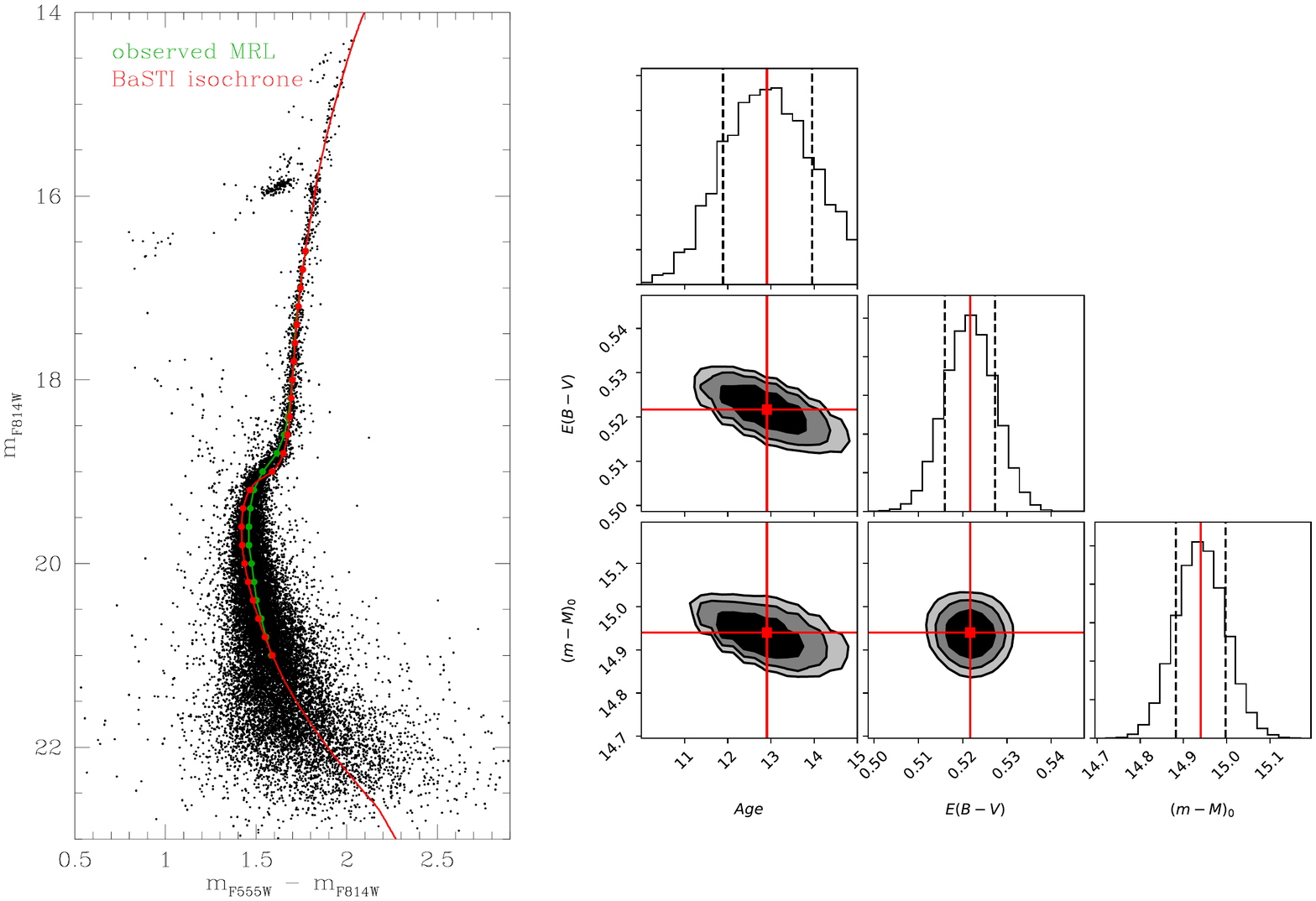}
	\caption{Similar to Figure \ref{fig:6569fig12} but for the BaSTI models. The best-fit parameter values are presented in Table \ref{6569_tab1}, right column.\label{fig:6569fig14}}
\end{figure}

The results obtained in terms of age, distance modulus and color excess are shown in Figures \ref{fig:6569fig12} - \ref{fig:6569fig14} for the three adopted sets of theoretical models.

In each figure the left-hand panel shows the observed optical CMD with the fiducial MRL drawn in green and the best-fit isochrone plotted as a red line. In all cases, the best-fit model reproduces reasonably well the cluster fiducial line along all of the evolutionary sequences in the CMD. We also verified that the lower envelope of the distribution of HB stars is also nicely reproduced up to bluer colors when the related ZAHB models are considered (\cite{pietrinferni2004}, VandenBerg (2018), private communication). The reduced (per degree of freedom) $\chi^{2}$ values (that are obtained from the color values marked with green and red dots in the figures) are 0.15, 0.20 and 0.48 for the DSED, VR and BaSTI isochrones, respectively. The one- and two-dimensional posterior probabilities for all of the parameter combinations are presented in the right-hand panel of each figure as corner plots. Reddening and distance are not mutually correlated, while there is a strong degeneracy of these two parameters with age.

%\texttt{DSED} isochrones: Figure \ref{fig:6569fig12}, {\it left panel} shows the best-fit DSED isochrone, superimposed on the ($m_{F814W}, m_{F555W}-m_{F814}$) CMD of NGC 6569. The fiducial line of the cluster is shown in dark green, for comparison. The best-fit solution is obtained for the following values: Age = $12.27^{+0.84}_{-0.74}$ Gyr, E(B-V) = 0.53 $\pm$ 0.01 mag and $(m-M)_{0}$ = $14.99^{+0.04}_{-0.04}$ mag (see also Table \ref{6569_tab1}, left column). The overall fit of the observed CMD is highly satisfactory, corresponding to a $\chi^{2}$ value of 3.94. The correlation between parameters pairs is presented in the right panel of Figure \ref{fig:6569fig12} as corner plots. Reddening and distance do not correlate each other but there is a strong degeneracy of these two parameters with the age.

%\texttt{VR} isochrones: The best-fit solution for the VR isochrones is shown in Figure \ref{fig:6569fig13}, {\it left panel}, where the color code is the same as in Figure \ref{fig:6569fig12}. This isochrone reproduces almost perfectly the fiducial line along all of the sequences in the CMD, with a $\chi^{2}$ value of 5.20. From the VR isochrones we find Age = $12.83^{+0.92}_{-0.85}$ Gyr, E(B-V) = 0.53 $\pm$ 0.01 mag and $(m-M)_{0}$ = $14.98^{+0.05}_{-0.04}$ mag (see also Table \ref{6569_tab1}, middle column). The one- and two-dimensional posterior probability for all of the parameter combinations are presented in Figure \ref{fig:6569fig13}, right panels.

%\texttt{BaSTI} isochrones: The result of the isochrone fitting for the BaSTI set of models is shown in Figure \ref{fig:6569fig14} and provides the following parameters: Age = $12.63^{+1.05}_{-0.94}$ Gyr, E(B-V) = 0.52 $\pm$ 0.01 mag and $(m-M)_{0}$ = $14.97^{+0.04}_{-0.04}$ mag (see also Table \ref{6569_tab1}, right column). This best-fit solution, which corresponds to $\chi^{2}$ = 10.67, very well reproduces our data over all the observed magnitude range. As can be seen, this BaSTI isochrone is also able to reproduce the observed magnitude of the RGB-bump of the cluster. The correlation among age, reddening and distance is shown in Figure \ref{fig:6569fig14}, right panels for BaSTI models, where the best-fit parameter values are marked in red.
%It is worth to point out that the BaSTI models have been computed by neglecting the effect of atomic diffusion. This introduces an offset in the age scale based on the three distinct model sets: the inclusion of the diffusion reduces the age by about 0.9 Gyr at the metallicity of the sample GCs (\citealt{cassisi1998,cassisi1999} and references therein). Therefore, the BaSTI-based age should be reduced by 0.9 Gyr, thus becoming Age = $12.19^{+1.06}_{-1.05}$ Gyr.

The best-fit values and their 1$\sigma$ uncertainties obtained for $E(B-V)$, $(m-M)_0$ and age from the three sets of models are listed in Table 1. They are all consistent within the errors. NGC 6569 turns out to be a quite old cluster, with an age of about 12.8 Gyr, and a typical age uncertainty of about 0.8 - 1.0 Gyr. Differences in the best-fit age are somewhat expected when different models, which adopt slightly different solar abundances, opacities, reaction rates, efficiency of atomic diffusion etc., are compared. We also note that an uncertainty of $\sim$ $\pm$ 0.1-0.2 Gyr has to be added to the overall error budget when the effect of a slightly different metallicity (by $\pm$ 0.1 dex) for the cluster is considered.  
%The color excess and distance modulus corresponding to the best-fit isochrones are of about $E(B-V)$ = 0.52 - 0.53, and $(m-M)_{0}$ = 15.02 - 15.04. %The $(m-M)_{0}$ estimates, within the uncertainties, agree with those derived in Section 6.   %In particular, BaSTI and VR isochrones have been computed by neglecting the effect of atomic diffusion, which has an important impact in modifying the MS-TO luminosity vs age relation. For this reason we need to reduce the best-fit age for both BaSTI and VR by 0.9 Gyrs (e.g. \citealp{kerber2018}), before comparing the results.
\begin{table}[ht!]
	\begin{center}
		\caption{Best-fit parameter values for the DSED, VR and BaSTI models.}\label{6569_tab1}
		\begin{tabular}{ c|c|c|c } 
			\hline
			&	DSED isochrone & VR isochrone & BaSTI isochrone \\
			\hline
			$E(B-V)$ [mag] & 0.53 $\pm$ 0.01 & 0.525 $\pm$ 0.010 & 0.52 $\pm$ 0.01 \\ 
			$(m-M)_{0}$ [mag]& $14.96^{+0.06}_{-0.06}$ & $14.95^{+0.06}_{-0.06}$ & $14.94^{+0.06}_{-0.06}$ \\ 
			Age [Gyr] & $12.51^{+0.88}_{-0.82}$ & $13.13^{+0.94}_{-0.90}$ & $12.91^{+1.05}_{-1.01}$ \\ 	
			\hline
		\end{tabular}
	\end{center}
\end{table} 

\subsection{Comparing NGC 6569 with other bulge GCs}
\begin{figure}
	\figurenum{16}
	\plotone{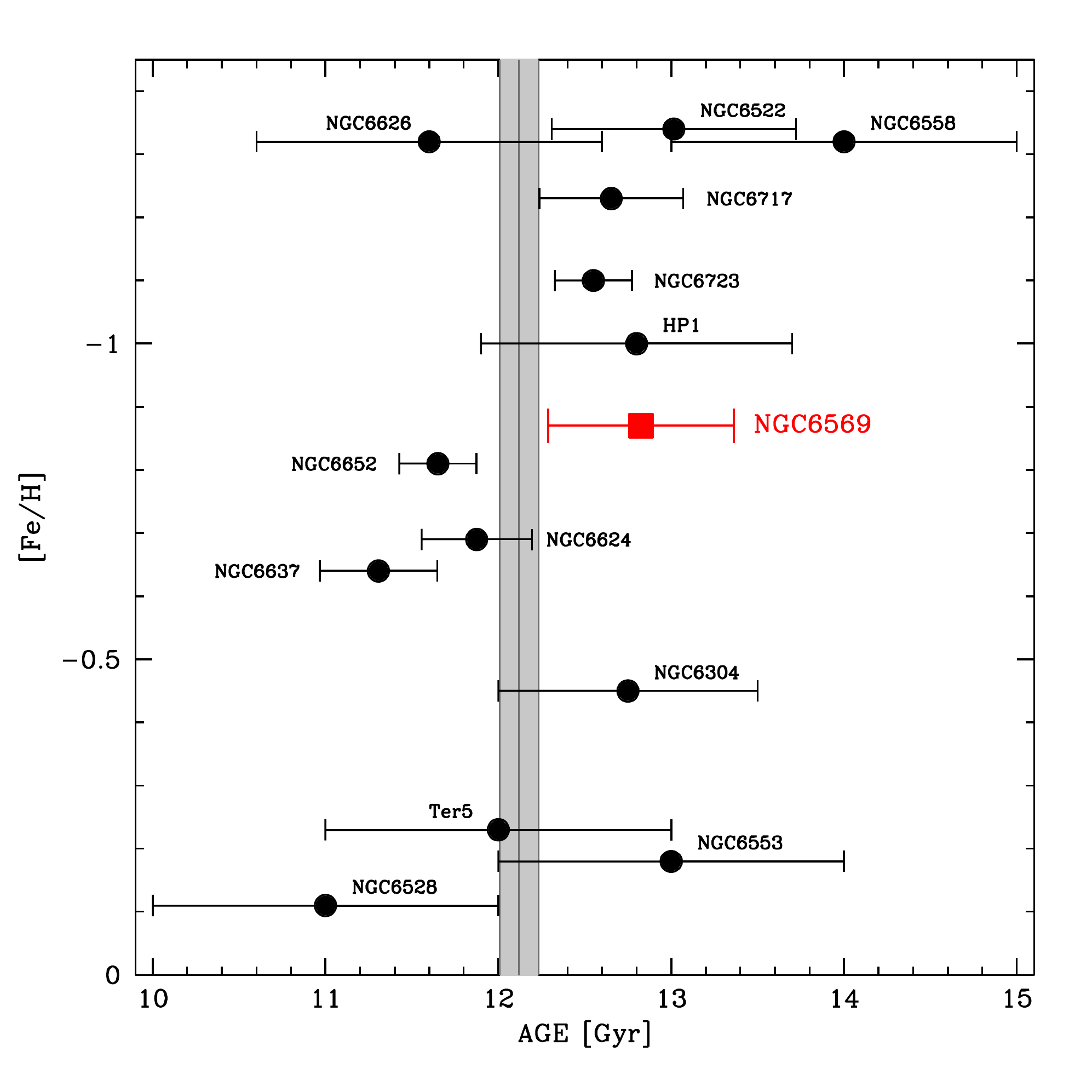}
	\caption{The absolute age of NGC 6569 (red square) derived in this work as weighted mean of the determinations given by the three sets of models listed in Table \ref{6569_tab1} is compared to that obtained for other bulge GCs in the literature (black circles; see references in Section 7.3), as a function of their metallicity. Different age estimates for a given system have been averaged in order to have a single value for each cluster. The vertical grey strip marks the weighted mean of the entire sample and its 1$\sigma$ uncertainty.\label{fig:6569fig15}}
\end{figure}

In this section we compare the age estimates obtained in the present work for NGC 6569 with those of other bulge GCs\footnote{The age estimates derived for different clusters come from slightly different approaches and/or assumptions, so that the direct comparison may be affected by some systematic uncertainties.}. Unfortunately up to now, only a dozen of bulge GCs have accurate age measurements, mainly because of their strong foreground extinction. The only cases are: {\it NGC 6304} \citep{dotter2010}; {\it NGC 6637, NGC 6652, NGC 6723} \citep{vandenberg2013,dotter2010}; {\it NGC 6522, NGC 6626} \citep{kerber2018}; {\it NGC 6624} \citep{saracino2016,vandenberg2013,dotter2010}; {\it NGC 6553} \citep{zoccali2001}; {\it NGC 6528} \citep{calamida2014}; {\it NGC 6558} \citep{barbuy2007}; {\it Terzan 5} (\citealt{ferraro2016}; here we take into account only the old, dominant component of the system) and {\it HP1} \citep{kerber2019}.
The current situation is summarized in Figure \ref{fig:6569fig15}, where we plot the age estimates currently available for bulge GCs, as a function of their metallicity (from \citealp{harris1996} with only two exceptions: NGC 6569, \citealp{johnson2018} and NGC 6624, \citealp{valenti2011}). For those systems for which more than one age value is quoted in the literature, we plot the weighted means and relative errors. All bulge GCs are old, with ages in excess of $\approx$ 11 Gyr, independently of their metallicity. The weighted mean age of the entire sample is 12.1 $\pm$ 0.1 Gyr, as marked by the vertical grey strip in Figure \ref{fig:6569fig15}. 
Such an old age of the bulge GC system is consistent with that of bulge stars observed in different fields (\citealp{zoccali2003}, \citealp{clarkson2011}, \citealp{valenti2013}), confirming that the stellar populations in the innermost regions of the Milky Way were born at the earliest epochs of the Galaxy's formation.
\section{Summary and Conclusions}
By using the high-resolution imaging cameras WFC3 on board {\it HST} in the optical and GeMS/GSAOI at GEMINI in the NIR, 
we obtained deep and accurate photometry in different filter combinations of the bulge GC NGC 6569. 
The derived CMDs span a range of about 8 magnitudes, allowing to identify all the well known evolutionary sequences, 
from the HB level down to about two magnitudes below the MS-TO.

For the first time we determined relative PMs for about 20000 individual stars in this cluster, sampling from the bright RGB to the faint MS. 
To this end, we used GEMINI and {\it HST} observations as first- and second-epoch data, respectively. 

We also determined the first differential reddening map in the direction of the system, 
finding that the color excess $E(B-V)$ varies by $\sim$0.12 magnitudes across the sampled FOV and it has a clumpy distribution. 
By using the observed position of the RGB-bump and the average V-band magnitude of a sample of 25 RR Lyrae stars, 
we find that NGC 6569 has a distance modulus $(m-M)_0$ = $15.03 \pm 0.08$, corresponding to $10.1 \pm 0.2$ kpc from the Sun.

Finally, by taking advantage of our high-resolution PM-cleaned and differential reddening-corrected 
CMD we obtained the first accurate estimate of the absolute age of NGC 6569, 
The isochrone fitting method, 
based on a $\chi^{2}$ statistics, has been used to compare three different sets of theoretical models (DSED, VR and BaSTI) 
with the observed ($m_{F814W}, m_{F555W}-m_{F814W}$) CMD of the cluster. By adopting as input ``guess'' values the 
distance modulus determined in this work and the reddening quoted in the literature (0.53, from Ortolani et al. 2001), 
we find that NGC 6569 has an absolute age of $12.8 \pm 1.0$ Gyr. 
We also confirmed the presence of a prominent, single red clump and of some blue HB stars. 

The promising results obtained in this study confirm that ground-based NIR imaging assisted by 
MCAO as provided by GeMS/GSAOI can be efficiently used for  
deep and accurate photometric and astrometric analysis of dense stellar fields in the central region of the Galaxy 
severely affected by extinction, 
once the geometric distortions of the detectors are properly corrected.

\section{Acknowledgements}
We thank the anonymous referee for the careful reading of the paper and for helping us to improve the presentation of the work. 
S.S. is grateful to Don Vandenberg for computing and sharing the Victoria-Regina ZAHB models used in the paper for comparison.  
D.G. gratefully acknowledges support from the Chilean Centro de Excelencia en Astrof\'isica
y Tecnolog\'ias Afines (CATA) BASAL grant AFB-170002.
D.G. also acknowledges financial support from the Direcci\'on de Investigaci\'on y Desarrollo de
la Universidad de La Serena through the Programa de Incentivo a la Investigaci\'on de
Acad\'emicos (PIA-DIDULS).

\software{\texttt{emcee} \citep{foremanmackey2013}, \texttt{CataXcorr}, \texttt{IRAF} \citep{tody1986,tody1993}, \texttt{DAOPHOTIV} \citep{stetson1987,stetson1994}, \texttt{ALLFRAME} \citep{stetson1994}}




\listofchanges

\end{document}